\documentclass[twoside,12pt]{article}
\usepackage{graphicx}
\usepackage{amssymb}
\usepackage{amsmath}
\usepackage{gensymb}
\usepackage{dcolumn}
\usepackage{bm}

\usepackage[labelfont={normalsize},subrefformat=parens]{subfig}

\newcommand{\be}{\begin{equation}}
\newcommand{\ee}{\end{equation}}
\newcommand{\vlowk}{V_{{\rm low}\,k}}
\newcommand{\mev}{\, \text{MeV}}

\newcommand{\pp}{{\bf p}}
\newcommand{\qq}{{\bf q}}
\newcommand{\rr}{{\bf r}}
\newcommand{\kk}{{\bf k}}
\newcommand{\jj}{{\bf j}}
\newcommand{\vv}{{\bf v}}

\oddsidemargin
\evensidemargin
\begin{document}

\begin{center}
{\large Pairing and superfluidity of nucleons in neutron stars}
\end{center}

\begin{center}
A.\ Gezerlis,$^{1,2,3}$ C.\ J.\ Pethick,$^{4,5}$ and A.\ Schwenk$^{2,3}$\\
\end{center}
$^1$Department of Physics, University of Guelph, Guelph,\\
Ontario, N1G 2W1, Canada\\
$^2$Institut f\"ur Kernphysik, Technische Universit\"at Darmstadt,\\
D-64289 Darmstadt, Germany \\
$^3$ExtreMe Matter Institute EMMI, GSI Helmholtzzentrum f\"ur\\
Schwerionenforschung GmbH, D-64291 Darmstadt, Germany\\
$^4$The Niels Bohr International Academy, The Niels Bohr Institute, 
University of Copenhagen, Blegdamsvej 17, DK-2100 Copenhagen \O, Denmark\\
$^5$NORDITA, KTH Royal Institute of Technology and Stockholm University, 
Roslagstullsbacken 23, SE-106 91 Stockholm, Sweden
\begin{center}
\today
\end{center}
\begin{center}
{\it ``Our life is frittered away by detail. \ldots Simplify, simplify.''\\
Henry David Thoreau, Walden.}
\end{center}

\begin{abstract}

\noindent
We survey the current status of understanding of pairing and
superfluidity of neutrons and protons in neutron stars from a
theoretical perspective, with emphasis on basic physical properties.
During the past two decades, the blossoming of the field of ultracold
atomic gases and the development of quantum Monte Carlo methods for
solving the many-body problem have been two important sources of
inspiration, and we shall describe how these have given insight into
neutron pairing gaps. The equilibrium properties and collective
oscillations of the inner crust of neutron stars, where neutrons
paired in a $^1$S$_0$ state coexist with a lattice of neutron-rich
nuclei, are also described. While pairing gaps are well understood at
densities less than one tenth of the nuclear saturation density,
significant uncertainties exist at higher densities due to the
complicated nature of nucleon-nucleon interactions, the difficulty of
solving the many-body problem under these conditions, and the
increasing importance of many-nucleon interactions. We also touch more
briefly on the subject of pairing of neutrons in other angular
momentum states, specifically the $^3$P$_2$ state, as well as pairing
of protons.

\end{abstract}


\tableofcontents


\newpage
 
\section{Introduction}
\label{sec:intro}
\subsection{Preamble and history}
In neutron stars one finds reservoirs of high-density fermions that are among the largest in the Universe and, because of the strong nucleon--nucleon interaction, a number of different phases can occur.  Understanding properties of these phases is necessary to interpret observations of neutron stars, which are becoming increasingly more detailed.  Temperatures in the interiors of neutron stars fall below a billion degrees Kelvin less than about one year after the birth of the star.  Such temperatures may appear high, but they are low compared with the characteristic  energies such as the Fermi energy, which in matter at nuclear density are typically $\sim10$ --$100$ MeV, corresponding to temperatures of order $10^{11}$--$10^{12}$ K.  Thus the effects of quantum degeneracy are important.

This chapter is devoted to pairing of neutrons and of protons in neutron stars at densities of order the saturation density of nuclei and below. The primary focus is on basic physical effects, on connections to other physical systems, and on topics where there has been significant recent progress.  The bibliography is illustrative rather than exhaustive.  Applications to observed neutron star phenomena are considered in another contribution to this volume \cite{page_this volume}.   For earlier reviews we refer to Refs.~\cite{LombardoSchulze,Dean:2002zx}.

Immediately after the Bardeen-Cooper-Schrieffer (BCS) theory of 
superconductivity was proposed, Bohr, Mottelson, and Pines \cite{Bohr:1958} showed that the excitation energies of the lowest lying non-collective states of nuclei were significantly  larger than could be accounted for on the basis of an independent-particle model.  They proposed an analogy between the low-lying spectra of atomic nuclei and that of a superconducting metal and argued that pairing between nucleons could account for a number of features of nuclei.  This idea was quickly followed up and led to profound insights into properties of nuclei (see, e.g.,~Ref.~\cite{BrinkBroglia}).  One of the manifestations of pairing in nuclei is a reduction of the moment of inertia of the nucleus, which results in an increase in the spacing of rotational levels compared with what would be expected for a rigid body.  In an early paper on this subject, Migdal remarked in passing ``We note that the superfluidity of nuclear matter can lead to interesting macroscopic phenomena if stars with neutron cores exist.  Such a star would be in a superfluid state with a transition temperature corresponding to 1 MeV.'' \cite{migdal}.  As we shall describe in greater detail in subsequent sections, Migdal's comment was remarkably prescient.   A few years later, Ginzburg and Kirzhnits \cite{ginzburgkirzhnits, ginzburg}
estimated pairing gaps and pointed to a number of consequences of superfluidity in neutron stars. Properties of vortex lines in superfluid neutrons were considered by Baym et al.~\cite{bpp}, who also argued that superfluid protons would behave as a Type II superconductor, in which magnetic flux would penetrate the medium in the form of quantized flux lines.  

\begin{figure}
\begin{center}
\includegraphics*[width=0.7\columnwidth]{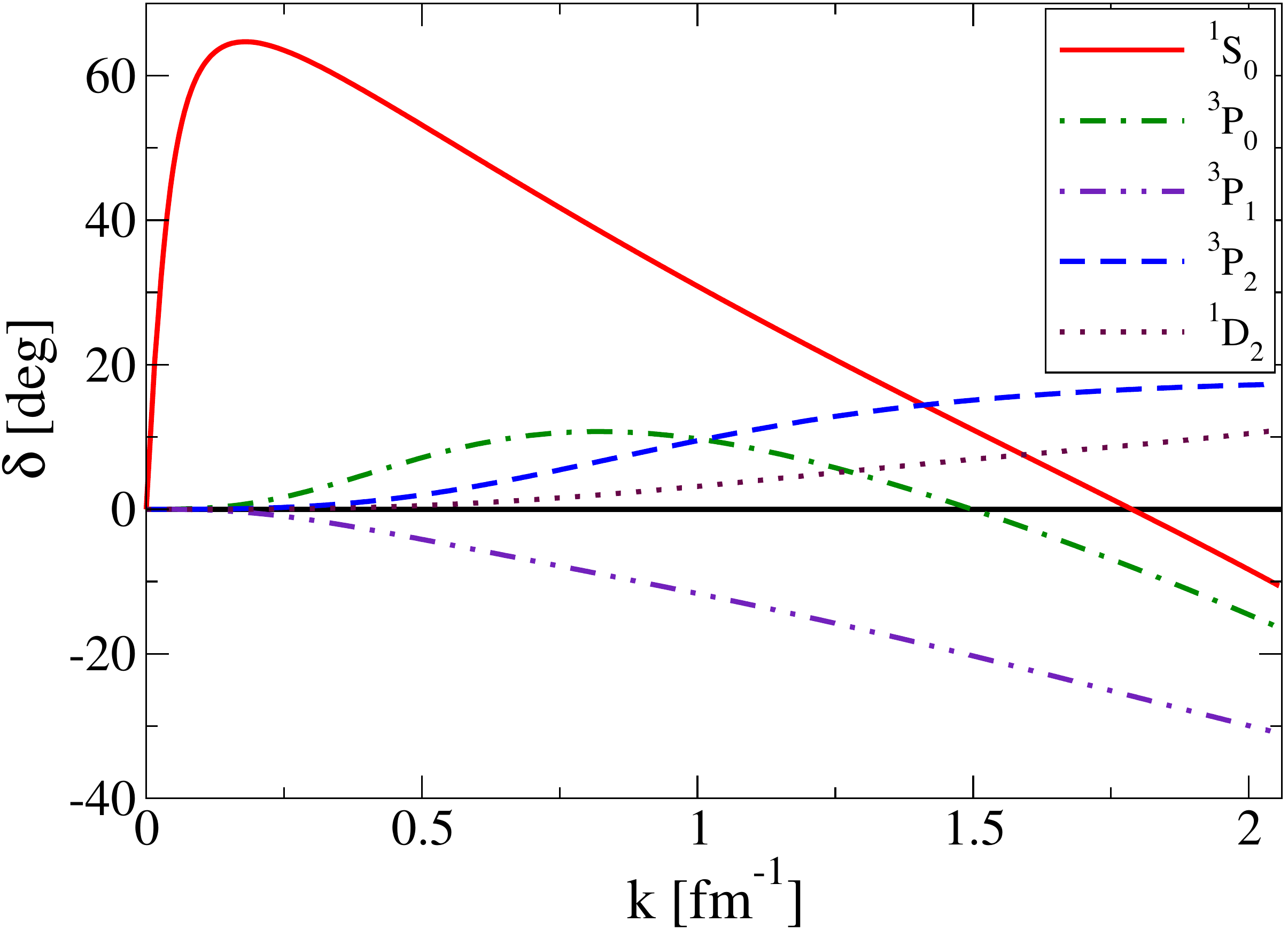}
\caption{Scattering phase shifts as a function of the wave number of a nucleon in the center of mass frame based on the Nijmegen 
partial-wave analysis of experimental data \cite{Stoks:1993}.   The results are for neutron–proton scattering but, due to the fact that isospin symmetry is only weakly broken, they provide a good approximation for neutron--neutron scattering.  Recall that a positive phase shift corresponds to an attractive interaction.} \label{fig:phaseshifts}
\end{center}
\end{figure}

Following the discovery of neutron stars, Hoffberg et al.\ \cite{hoffberg} calculated gaps for neutron matter within the BCS theory using a separable nucleon-nucleon interaction that had been fitted to two-nucleon scattering data.  Their calculations predicted a pairing gap for neutrons in the $^1$S$_0$ state\footnote{We use the standard spectroscopic notation $^{2S+1}$L$_J$ to specify the angular momentum state of the paired particles.  Here $S$ is the total spin of the two particles, $L$ is the orbital angular momentum of their relative motion, and $J$ is the total angular momentum.} that first rose with increasing density, reached a maximum of roughly 3 MeV at a density of about $n_s/10$, where $n_s \approx 0.16$ fm$^{-3}$ is the saturation density of nuclear matter with equal numbers of neutrons and protons, a density typical of the interiors of heavy nuclei.  With further increase in density, the gap dropped and vanished at a density somewhat below $n_s$. At that density it had already become favorable for neutrons to pair in the $^3$P$_2$ state, in which the pairs have unit orbital angular momentum, unit spin, and total angular momentum 2.   That the $^3$P$_2$ state has a more attractive interaction than the other $^3$P states is due to the fact that the spin--orbit interaction is attractive for nucleons.  This situation is to be contrasted with that in atomic physics, where the spin--orbit interaction is repulsive.  The $^3$P$_2$ gap increased to about 0.5 MeV at a density of around $2n_s$ and dropped at higher densities.  The qualitative behavior of the gaps may be understood in terms of the measured phase shifts for nucleon--nucleon interactions, which are displayed in Fig.\ \ref{fig:phaseshifts}.  A positive phase shift corresponds to an attractive interaction between neutrons, and therefore at low $k$, which corresponds to low Fermi momentum and low density, the most attractive channel is $^1$S$_0$, while at higher densities the interaction in the $^3$P$_2$ channel is more attractive.

In what we (in common with most of the literature in nuclear physics) refer to as the BCS approximation, the gap is calculated by solving the BCS equation with the free-space nucleon--nucleon interaction and for free particles in intermediate states in the scattering process: the effects of the neutron medium on the normal state excitations and the pairing interaction are neglected.\footnote{We make the definition of the phrase  ``BCS approximation'' explicit because, in the physics literature, there are a number of usages of the phrase, among which is the  BCS schematic model for the interaction, in which one assumes that the interaction matrix element is constant for a range of momentum states in the vicinity of the Fermi surface.} A variety of techniques have been used to include effects beyond the BCS approximation in calculations of  neutron pairing, and they typically predict a reduction of the pairing gap by a factor of 2 or more.   The earliest estimates of the effects of the medium were made with variational methods involving wave functions that include the effects of correlations \cite{clark} and with the use of methods inspired by diagrammatic perturbation theory \cite{Wambach:1993}.

In neutron stars, S-wave pairing is predicted to occur at subnuclear densities and the bulk properties of neutron matter are an important ingredient in understanding matter at such densities despite the fact that the neutrons coexist with a lattice of nuclei and an electron gas.  Two developments over the past two decades have been important for understanding this region.  One is the increasing power of quantum Monte Carlo methods to make accurate predictions based on the underlying interactions between individual nucleons.  The second is the experimental realization of atomic gases at ultralow temperatures, where quantum effects manifest themselves.

The remainder of this chapter is organized as follows: in Sec.\ 1.2 we describe the BCS theory and its application to the neutron liquid.  In Sec. 2 we describe pairing in a dilute Fermi gas and show that the reduction of the pairing interaction due to screening by the medium gives rise to a suppression of the pairing gap even in the limit of very low densities.   The analysis there provides valuable insight into the results of detailed many-body calculations for neutron matter.  In addition,  the physics of neutron matter is compared and contrasted with that of ultracold atomic gases.  Microscopic calculations of gaps are described in Sec.\ 3; besides the $^1$S$_0$ state in neutron matter, we consider the $^3$P$_2$ state, and pairing of protons in the  $^1$S$_0$ state.  The inner crust of neutron stars, where superfluid neutrons coexist with a crystal lattice of neutron-rich nuclei and an electron gas, is the subject of Sec.\ 4.  There we begin with calculations of static properties, before going on to describe a hydrodynamic approach to long-wavelength dynamics and collective modes, and the problem of determining the neutron superfluid density.   Section 5 contains concluding remarks.

\subsection{BCS theory}
\label{sec:BCS}

We begin by giving a brief overview of BCS theory mainly to establish notation.
For an unpolarized system of  spin-1/2 particles interacting via a two-body potential, 
the BCS wave function for the ground state includes pairing between particles in a spin-singlet state with zero orbital angular momentum or, equivalently, between particles with equal and opposite spin and momentum, and it has the form (see e.g. 
Refs. \cite{Degennes:1966,Tinkham:1996})
\begin{equation}
| \psi_{BCS} \rangle = \prod_{\bf k} (u_{{\bf k}} + v_{{\bf k}} c^{\dagger}_{{\bf k} \uparrow}
c^{\dagger}_{{-\bf k} \downarrow})|0\rangle~,
\label{eq:bcswave}
\end{equation}
where the operator $c^{\dagger}_{{\bf k} \sigma}$ creates a particle with wave vector  $\bf k$ and spin projection $\sigma=\uparrow, \downarrow$ and 
$c_{{\bf k} \sigma}$ destroys one.  The coefficients  $u_{\bf k}$ and  $v_{\bf k}$, which without loss of generality may be taken to be real and positive, satisfy the normalization condition 
\be
u_{{\bf k}}^2 + v_{{\bf k}}^2 = \,1.
\label{norm}
\ee
  For the wave function
 (\ref{eq:bcswave}) the probability of the single-particle state
${\bf k} \sigma$ being occupied is $v_{{\bf k}}^2$ and the probability that it is unoccupied is $u_{{\bf k}}^2 = 1 - v_{{\bf k}}^2$. 

The wave function (\ref{eq:bcswave}) is not an eigenstate of the  number of 
particles, ${\hat N}=\sum_{{\bf k}, \sigma}c^{\dagger}_{{\bf k} \sigma}c_{{\bf k} \sigma}$. Therefore it is convenient to work in the grand canonical ensemble, in which the average number of particles, $\bar{N}$, is fixed, the bar indicating an expectation value for the BCS wave function. The average particle number is given by
\begin{equation}
\bar{N} = \langle \hat{N} \rangle = \langle  \psi_{BCS} | \sum_{\bf k} \left ( c^{\dagger}_{{\bf k} \uparrow} c_{{\bf k} \uparrow} + c^{\dagger}_{{\bf k} \downarrow} c_{{\bf k} \downarrow} \right ) | \psi_{BCS} \rangle = 2 \sum_{\bf k}  v_{{\bf k}}^2~.
\label{eq:averageN}
\end{equation}
In the BCS approximation, one calculates the expectation value of the Hamiltonian $H$ for the wave function  (\ref{eq:bcswave}). Terms in the interaction energy that do not correspond to pairing correlations but rather to Hartree and Fock contributions, e.g., ones with factors of the form  $v_{\bf k}^2 v_{{\bf k}'}^2$, are neglected and the result is
\begin{equation}
\langle H - \mu \hat{N} \rangle  = 2 \sum\limits_{{\bf k}}  \xi({\bf k}) v_{{\bf k}}^2 + \sum\limits_{{\bf k k'}}  \langle {\bf k} | V | {\bf k'} \rangle u_{{\bf k}} v_{{\bf k}} u_{{\bf k'}} v_{{\bf k'}}~,
\label{eq:bcsintermhami}
\end{equation}
where $\xi({\bf k}) = \epsilon({\bf k})-\mu$, $\epsilon({\bf k}) = {\hbar^2k^2}/{2m}$ is the kinetic 
energy of a particle with wave vector ${\bf k}$, $\mu$ is the chemical potential, and $ \langle {\bf k} | V | {\bf k'} \rangle$ is the matrix element of the interaction Hamiltonian between a two-body state with a pair of particles with opposite spin and wave vectors $\pm {\bf k}$ and a similar state in which the particles have wave vectors $\pm {\bf k}'$. Minimizing this
expression subject to the normalization constraint (\ref{norm}) one obtains the so-called gap equation
\begin{equation}
\Delta({\bf k}) = -\sum_{{\bf k'}} \langle {\bf k} | V | {\bf k'} \rangle \frac{\Delta({\bf k'})}{2E({\bf k'})}~,
\label{deltageneraleq}
\end{equation}
where the gap is given by the expression
\be
\Delta({\bf k})=-\sum_{{\bf k}'}\langle {\bf k} | V | {\bf k'} \rangle u_{{\bf k}'}v_{{\bf k}'}.
\label{gap}
\ee
Here $E({\bf k}) = \sqrt{\xi({\bf k})^2+\Delta({\bf k})^2}$ 
is the quasiparticle excitation energy and in terms of this one finds
\be
u_{\bf k}^2=\frac12\left(1+\frac{\xi({\bf k})}{E({\bf k})} \right)\,\,\,{\rm and}\,\,\,v_{\bf k}^2=\frac12\left(1-\frac{\xi({\bf k})}{E({\bf k})} \right).
\ee
The chemical potential is found by solving the gap 
equation together with the equation for the average particle number,
\begin{equation}
\langle N \rangle = \sum_{{\bf k}} \left [ 1 - \frac{\xi({\bf k})}{E({\bf k})} \right ]~.
\label{particlegeneraleq}
\end{equation}

Equations (\ref{deltageneraleq}) and (\ref{particlegeneraleq}) can be decoupled when $\Delta/\mu \ll 1$, a condition that is not satisfied for 
the strongly paired systems discussed here (ultracold atomic gases with resonant interactions and neutrons at relatively low densities). Therefore, these two equations have to be solved self-consistently. The BCS approximation gives a good qualitative description of the pairing gap but, as we shall describe in following sections, there are quantitatively significant  effects due to correlations in the medium.

\begin{figure}
\begin{center}
\includegraphics*[width=0.7\columnwidth]{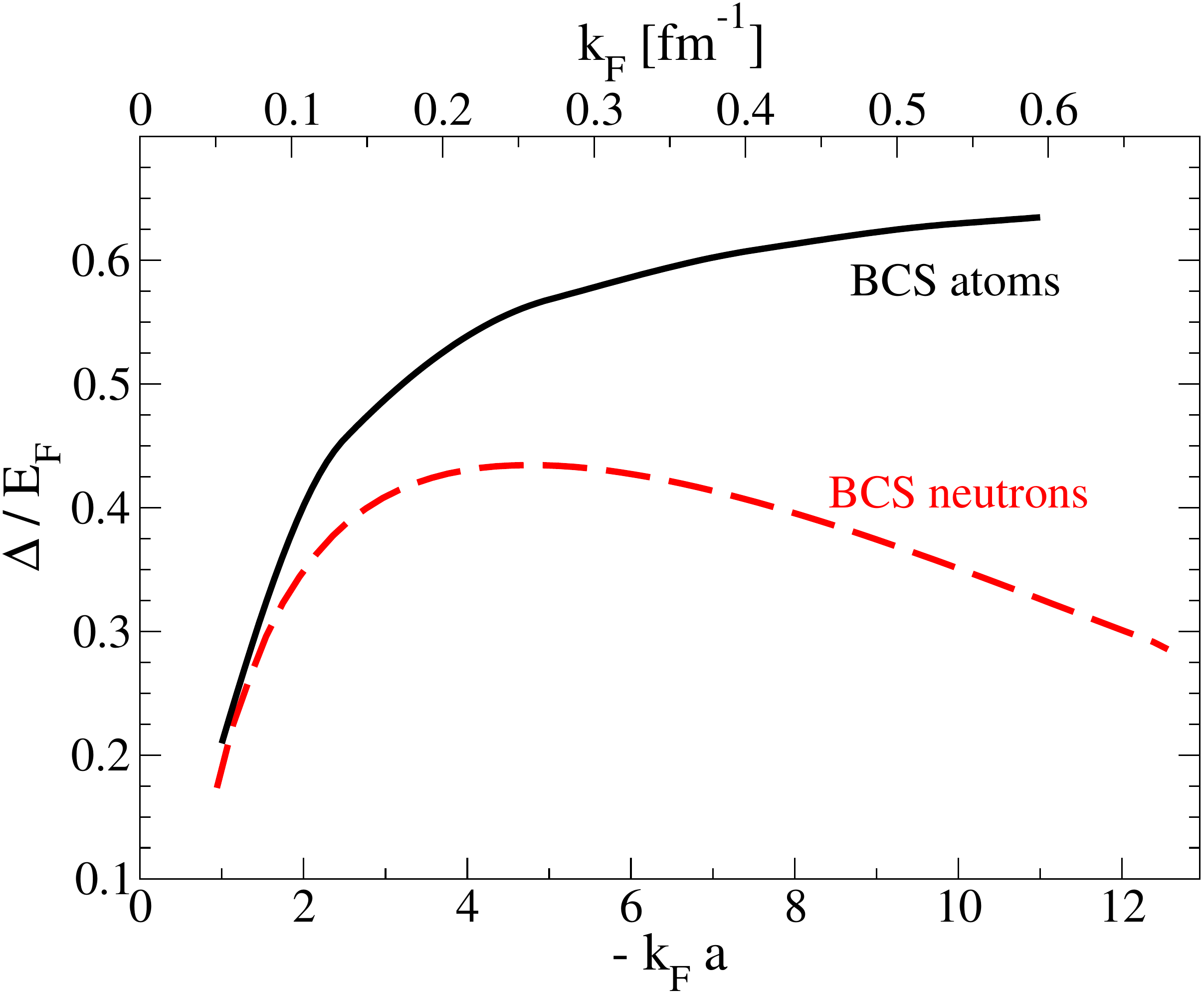}
\caption{Pairing gap in the BCS approximation in terms of the Fermi energy $E_F=\hbar^2k_F^2/2m$ versus $-k_F a$ for cold atoms and 
neutron matter (for details of the potentials used, see text). The upper scale is the Fermi wave number $k_F$ for neutron matter when the scattering length is taken to have its experimentally determined value 
$a = -18.5$ fm.  For $k_F |a|\lesssim 1$ the two results agree, 
which shows that, under these conditions, the gap is independent of the nonzero range of the neutron--neutron interaction.}
\label{fig:bcscomp}
\end{center}
\end{figure}

For pairing in the $^1\mbox{S}_0$ state, the gap is independent of the direction of $\kk$ and  Eq. (\ref{deltageneraleq}) takes the form
\begin{equation}
\Delta(k) = -\frac{1}{\pi} \int_0^{\infty} dk' ~k'^2 \frac{V(k,k')}{E(k')} \Delta(k')~,
\label{deltaconteq}
\end{equation}
where $V(k,k')$, the matrix element of the potential averaged over the angle between $\kk$ and $\kk'$, is given by
\begin{equation}
V(k,k') = \int_0^{\infty} dr ~r^2 j_{0}(k'r) V(r) j_{0}(kr)~,
\end{equation}
$j_{0}(kr)$ being the spherical Bessel function of zeroth order. Similarly, Eq.\ (\ref{particlegeneraleq}) may be written as an expression for the particle number density $n=\langle N \rangle/\Omega$, where $\Omega$ is the volume of the system:
\begin{equation}
n = \frac{1}{2\pi^2} \int_0^{\infty} dk ~k^2 \left(1 - \frac{\xi(k)}{E(k)}\right)~.
\label{particleconteq}
\end{equation}
These new equations are one-dimensional, and thus easier to solve numerically.  We have solved Eq.\ (\ref{deltaconteq}) together with Eq.\ (\ref{particleconteq}) for 
the $^1\mbox{S}_0$ channel of the Argonne $v_{18}$  potential \cite{Wiringa:1995},
which contains a strong short-range repulsion. This calculation is simplified by transforming the problem into a quasilinear one, as described in Ref.\ \cite{Khodel:1996}. 
We have also solved these equations 
for a potential $V(r)=B \,{\rm sech}^2(r/d)$ with the strength $B$ and range parameter $d$ tuned so that it describes cold atomic systems, for which the effective range
is much shorter than the interparticle spacing (see Fig. \ref{fig:bcscomp}) \cite{Gezerlis2008}. The main difference between the two
is that for cold atoms the gap in terms of the Fermi energy $E_F = \hbar^2 k_F^2 /2m$ continues to rise with increasing $k_F |a|$ and saturates at a finite, nonzero value for $k_F|a| \to \infty$ while for neutrons the gap drops to zero at a finite density.  This latter effect is
due to the fact that, as shown in Fig.\ 1, at large momenta the $^1\mbox{S}_0$ phase shift  becomes negative, corresponding to a repulsive interaction.

\section{Dilute neutron gas}
\label{sec:dilute}
At low energies, the effective interaction between two particles is determined by the S-wave scattering length, $a$.  For two neutrons in a singlet spin state, the scattering length is $-18.5$ fm, which is large compared with the range of nuclear interactions, $\sim 1$ fm.    For densities much less than $1/a^3 \sim 10^{-4}$ fm$^{-3}$, the leading interaction contributions to the properties of the system may be calculated in terms of an effective interaction of the form
\be
U_0=\frac{4\pi \hbar^2 a}m
\ee
in momentum space, which corresponds to a delta function in coordinate space.  The study of dilute quantum gases, which dates back to the period when many-body theory was in rapid development in the 1950s and 1960s, has experienced a renaissance following the experimental realization of such systems with ultracold atoms \cite{ColdFermions}. This has led to considerable insight into the properties of neutron matter. 
The condition for a gas to be dilute is that the interparticle spacing, $r_s$, be large compared with the magnitude of the scattering length $a$ of the particles or, since the Fermi wave number $k_F$ is proportional to $1/r_s$, this condition is equivalent to  $k_F|a|\ll 1$.    In a very important paper, which was overlooked for many years, Gor'kov and Melik-Barkhudarov studied the transition temperature and pairing gap for such a system~\cite{Gor'kov} and we begin by explaining their calculation in physical terms~\cite{HeiselbergPSV}.  

In standard BCS theory (Sec.\ \ref{sec:BCS}), one takes into account repeated scattering between a pair of particles via the free-space two-body interaction.  If the interaction is eliminated in favor of the scattering length, one finds for a dilute Fermi gas with two spin components of equal mass and equal number densities for the pairing gap $\Delta_{\rm BCS}$ at zero temperature the result
\be
\Delta_{\rm BCS}=\frac{8}{e^2}E_Fe^{-1/N(E_F)|U_0|}\approx 1.083 E_Fe^{-1/N(E_F)|U_0|},
\label{BCSgap}
\ee  
where the scattering length is taken to be negative.  The quantity $N(E_F)=mk_F/(2\pi^2\hbar^2)$ is the density of states of a single spin species at the Fermi surface. Somewhat surprisingly, by systematically investigating the low-density limit, Gor'kov and 
Melik-Barkhudarov found 
\be
\Delta=\left(\frac{2}{e}\right)^{7/3}E_Fe^{-1/N(E_F)|U_0|} \approx 0.489E_Fe^{-1/N(E_F)|U_0|},
\label{GMBgap}
\ee
a factor $(4e)^{1/3}\approx 2.22$ smaller than the BCS value.  The difference between the two results is due to the influence of the medium on the interaction between pairs of particles, as we shall now explain.

\subsection{Induced interactions}
\label{sec:inducedinteractions}

In a dilute Fermi gas with a short-range interaction there is very little interaction between particles of the same spin because the Pauli principle hinders particles in the same spin state from coming close together.  
\begin{figure}
\begin{center}
\includegraphics*[width=0.5\columnwidth]{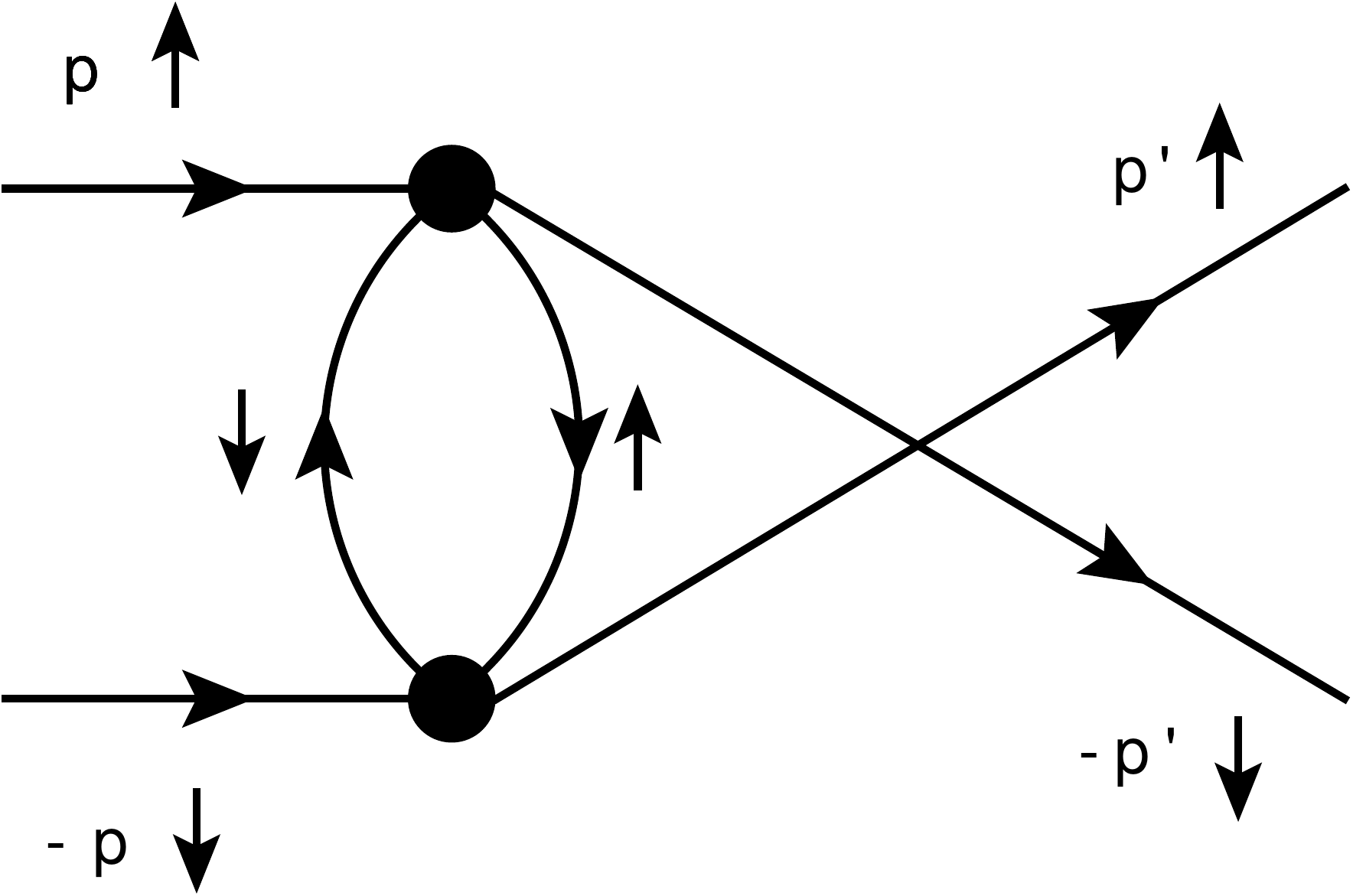}
\caption{Diagrammatic representation of the simplest modification of the effective interaction due to the surrounding medium.}
\label{BubbleExchangeCrossed2}
\end{center}
\end{figure}
Consequently, the interaction may be taken to operate only between particles of opposite spin, in which case the simplest modification of the interaction is the screening process shown in Fig. \ref{BubbleExchangeCrossed2}, and its contribution for zero energy transfer between the particles is
\be
U_{\rm ind}=U_0^2L(|\pp +\pp'|),
\ee
where 
\be
L(q)=N(E_F)\left[ \frac12+\frac{(1-w^2)}{4w}\ln\left|\frac{1+w}{1-w}  \right|   \right] ,
\ee
with $w=q/2p_F$, is the static Lindhard function familiar from the theory of screening in the electron gas and the Fermi momentum is given by $p_F=\hbar k_F$.
For scattering of two particles on the Fermi surface, $q$ is given by $2p_F \cos(\theta/2)$ and therefore when averaged over the angle $\theta$ between $\pp$ and $\pp'$ on the Fermi surface one finds
\be
\langle U_{\rm ind}\rangle=  N(E_F)\frac{U_0^2}3(1 +2\ln 2)=2N(E_F)U_0^2\ln[(4e)^{1/3}] \,,
\ee
where $\langle \ldots \rangle$ denotes the average over the Fermi surface.
On replacing the interaction $U_0$ in the BCS expression for the gap, Eq.\ (\ref{BCSgap}), by $U_0$ plus the average of the induced interaction one finds to lowest order in the small parameter $N(E_F)U_0=(2/\pi)k_F a$ that the gap is reduced by a factor $(4e)^{1/3}$, as we mentioned in Eq.~(\ref{GMBgap}) above. 

\begin{figure}
\begin{center}
\includegraphics*[width=0.5\columnwidth]{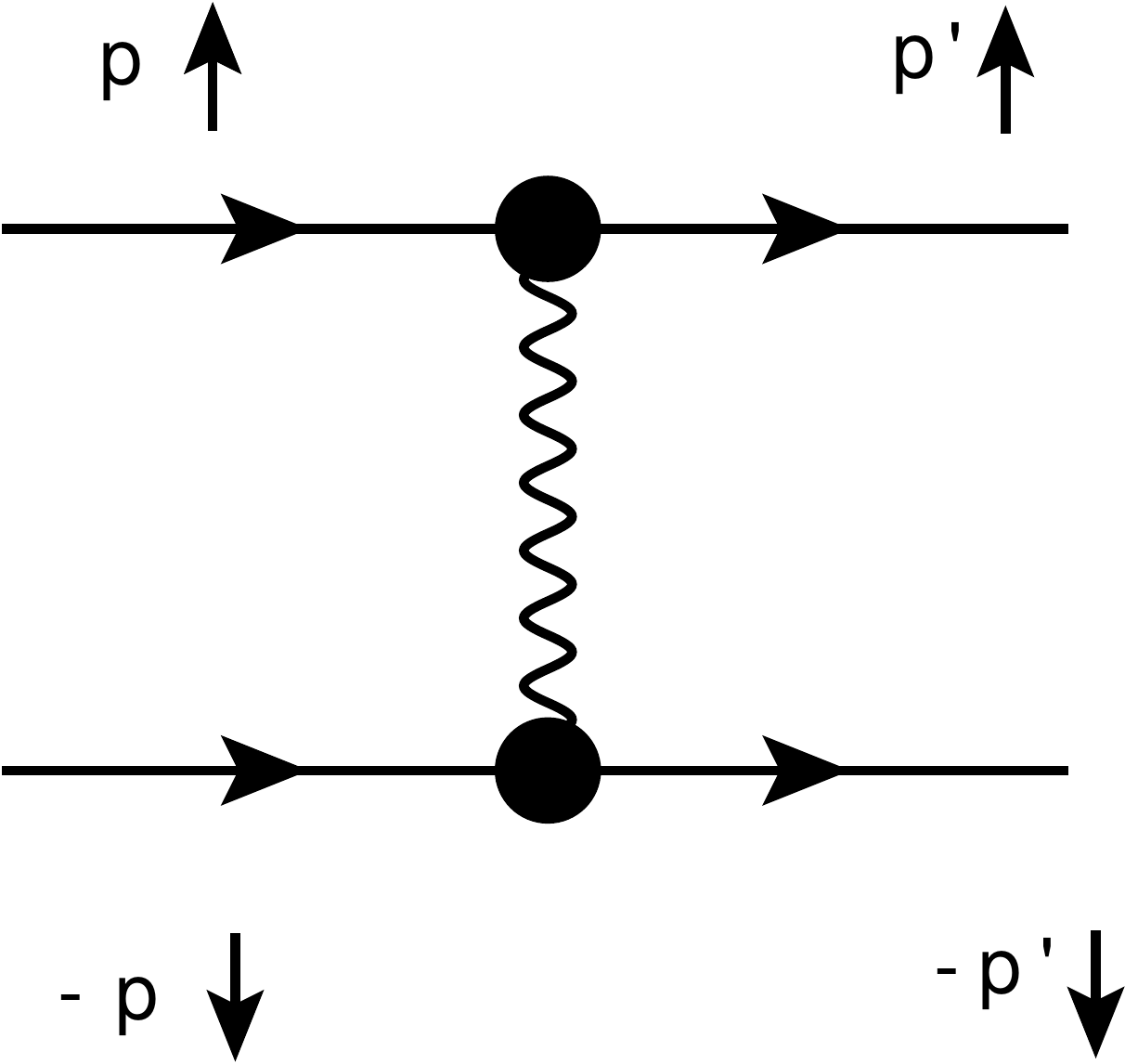}
\caption{Diagrammatic representation of the induced interaction due to exchange of a density fluctuation in the surrounding medium.}
\label{Fig:PhononExchange2}
\end{center}
\end{figure}

\begin{figure}
\begin{center}
\includegraphics*[width=0.5\columnwidth]{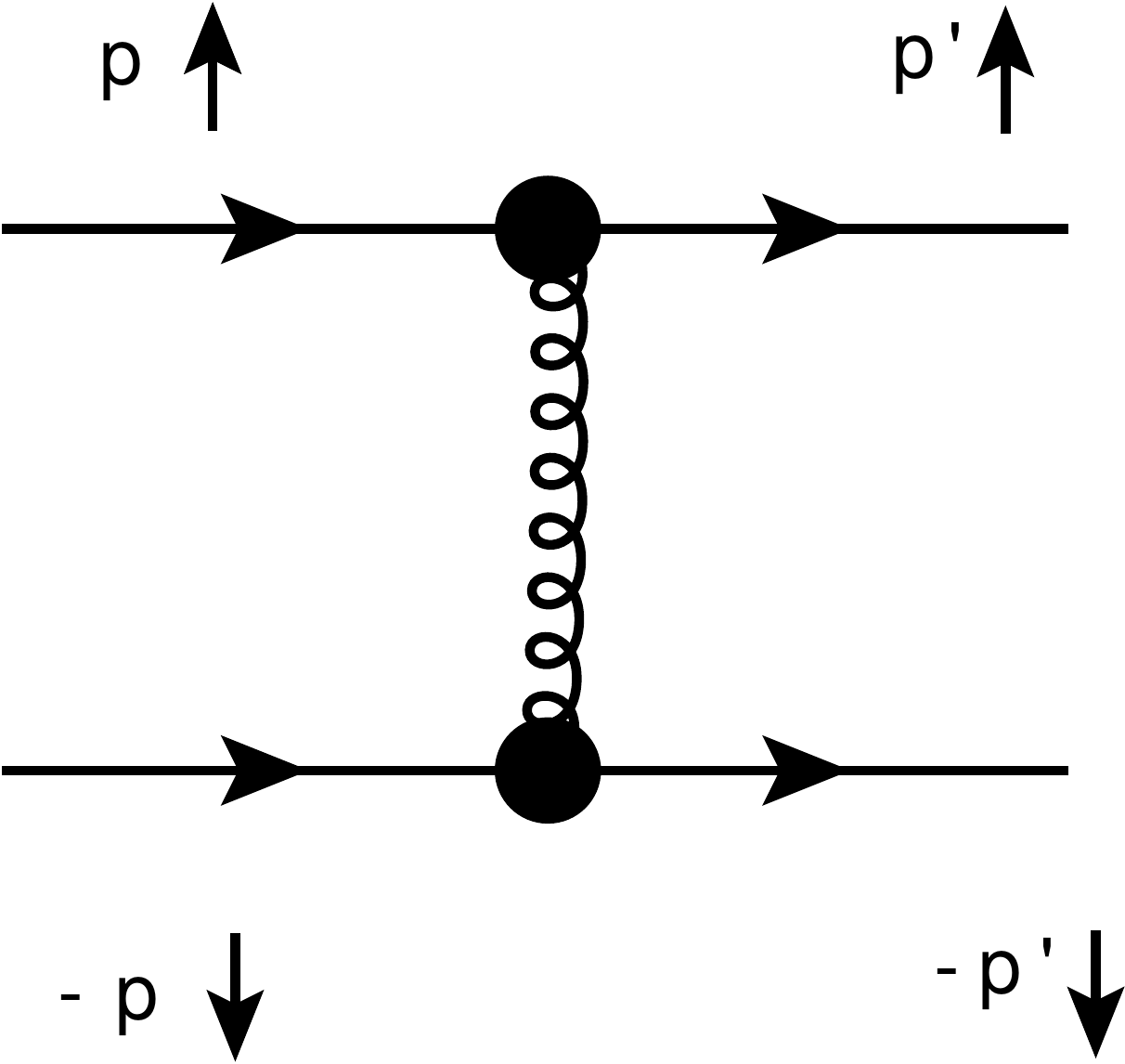}
\caption{Diagrammatic representation of the induced interaction due to exchange of a spin fluctuation with spin projection $m_S=0$ along the quantization axis.}
\label{Fig:SpinExchange2}
\end{center}
\end{figure}
\begin{figure}
\begin{center}
\includegraphics*[width=0.7\columnwidth]{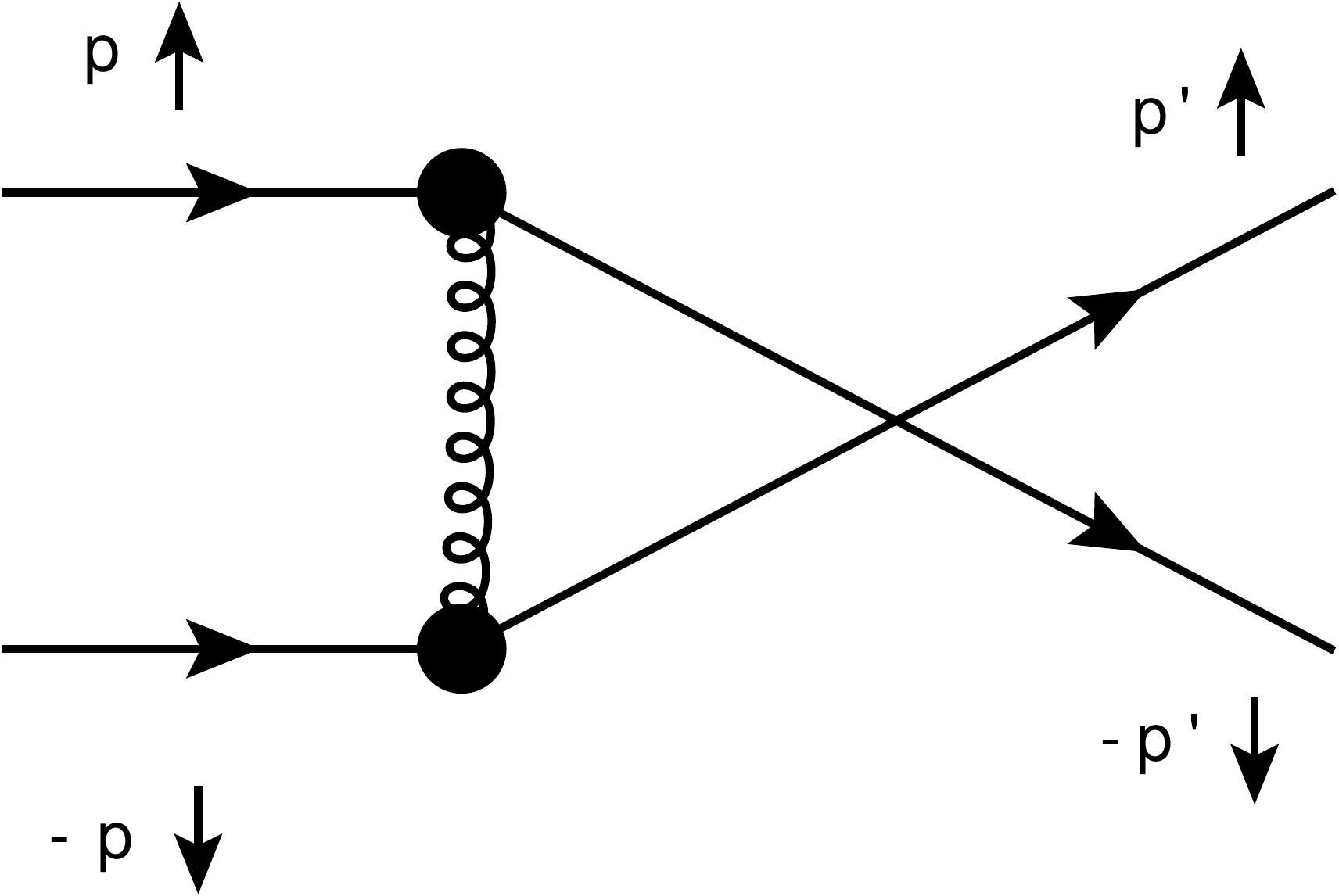}
\caption{Diagrammatic representation of the induced interaction due to exchange of a spin fluctuation with spin projection $m_S=\pm1$ along the quantization axis.}
\label{Fig:SpinExchangeCrossed2}
\end{center}
\end{figure}

Further insight may be obtained by analyzing the induced interaction in terms of exchange of excitations in the medium.  A short-range interaction between fermions is described by the Hamiltonian 
\be
H_{\rm int}=U_0\int d^3r\, n_\uparrow(\rr) n_\downarrow(\rr)=\frac{U_0}{4}\int d^3r\, [n(\rr)^2- n_s(\rr)^2],
\label{contactinteraction}
\ee
where $n_\uparrow(\rr)={\hat\psi}^\dagger_\uparrow(\rr) {\hat\psi}_\uparrow(\rr)$ and $ n_\downarrow(\rr)={\hat\psi}^\dagger_\downarrow(\rr) {\hat\psi}_\downarrow(\rr)$. The total number density is given by $n(\rr)=n_\uparrow(\rr)+n_\downarrow(\rr)$, and the spin density by $n_s(\rr)=n_\uparrow(\rr)-n_\downarrow(\rr)$.  Here ${\hat\psi}^\dagger_\sigma(\rr)$ is the operator that creates a particle of spin projection $\sigma$ at point $\rr$ and ${\hat\psi}_\sigma(\rr)$ is the corresponding annihilation operator.  A repulsive interaction between particles with opposite spin therefore leads to a repulsive interaction between particle densities and an attractive one between spin densities.

The existence of the medium changes the interaction between two particles: in addition to the direct interaction between two particles, there are contributions due to exchange of excitations in the medium.  One of the best known examples of this is the attractive interaction between electrons due to exchange of lattice phonons, which we show schematically in Fig.\ \ref{Fig:PhononExchange2}.    The wiggly line represents a density fluctuation in the medium, which in the case of lattice phonons is a well-defined mode of the system, but in a Fermi liquid corresponds to a collection of particle-hole pairs as well as any well-defined collective modes that may exist.  In an interacting Fermi system, spin fluctuations can also be exchanged, as illustrated in Fig.\ \ref{Fig:SpinExchange2}, where the curly line represents a spin fluctuation.  This term is repulsive, because particles with opposite spin couple to spin fluctuations with opposite signs.   In a dilute Fermi gas, the contribution from exchange of density fluctuations is exactly cancelled by that from spin fluctuations because, first, the density--density response function
(the wiggly line) and the spin-density--spin-density response functions are both equal to the Lindhard function $L$ and the coupling of particles to spin and density fluctuations are equal in magnitude (see Eq.\ (\ref{contactinteraction})).   In Fig.\ \ref{Fig:SpinExchange2} the spin fluctuation carries no net spin projection along the quantization axis ($m_S=0$).  However, it is also possible to exchange spin fluctuations with $m_S=\pm1$, as shown in Fig.\ \ref{Fig:SpinExchangeCrossed2}.  This term is also repulsive since, while the matrix elements of the spin-raising and spin-lowering operators at the vertices have the same sign, in contrast to matrix elements for the $m_S=0$ case, the interaction has the nature of an exchange term and therefore acquires an additional minus sign.  
In a more general model in which particles couple to fluctuations of  both number density and spin density with matrix elements $g_n$ and $g_s$ which are independent of momentum transfer, the contribution of induced interactions to the averaged pairing interaction from diagrams like those in Figs.\ \ref{Fig:PhononExchange2}, \ref{Fig:SpinExchange2} and \ref{Fig:SpinExchangeCrossed2} is
\be
\langle U_{\rm ind}\rangle =-g_n^2\langle\chi_n\rangle +3g_s^2\langle\chi_s\rangle,
\ee
where $\chi_n$ and $\chi_s$ are the static density and spin-density propagators (the negative of the density--density and spin-density--spin-density response functions).  The factor 3 in the spin-fluctuation term is due to the fact that the spin exchanged can lie in any of the three spatial  directions. That spin fluctuations tend to suppress S-wave superconductivity in metals has been known since the pioneering paper of Berk and Schrieffer \cite{berkschrieffer}, but the mechanism is quite general.   

In liquid $^3$He the interaction between atoms  induced by exchange of spin fluctuations is also responsible for the 
relative stability of the anisotropic superfluid phases, in which atoms are paired in P-wave states \cite{andersonbrinkman}.
In nuclear matter, the fermions have four spin-isospin states and therefore the above discussion for two internal states needs to be extended.
In the context of ultracold atomic gases, the effect of fluctuations in three-component systems has been investigated in Ref.\ \cite{martikainen} and these ideas could also be exploited to study, e.g., a proton impurity in an neutron gas with both spin states populated.

The effects of the medium on the pairing interaction are important even at very low densities because first, these contributions are of order $k_Fa$ times the lowest-order contribution and, secondly, the gap depends exponentially on the inverse of the pairing interaction.  The medium also affects the quasiparticle spectrum but this does not change the leading behavior of the gap on $k_Fa$ since corrections to the spectrum (as reflected in, e.g.,  the effective mass) are of order $(k_Fa)^2$. 

\subsection{Comparison with ultracold atomic gases}

One of the remarkable features of nucleon--nucleon interactions is that they are close to resonant at low energy.  This is reflected in the fact that the scattering lengths for nucleons  are much greater in magnitude than the range of the nucleon--nucleon interaction, $\sim 1$~fm.  For two neutrons in a singlet state, $a=-18.5$ fm.  Expressed in other terms, the neutron--neutron interaction is almost, but not quite, strong enough to create a bound state of two neutrons, a dineutron.  For a neutron and a proton, the scattering lengths are again large, with values $-23.7$~fm for the singlet state and  $+ 5.4$ fm for the triplet state.  The large, positive value for the triplet state is due to the interaction in that channel being sufficiently strong to form a bound state, the deuteron, which has a binding energy much smaller than the typical depth of nucleon--nucleon interactions.  As a consequence of the existence of the bound state,  the effective low-energy interaction is repulsive.    

Beginning in the 1990s, much progress was made in studying experimentally the properties of cold atomic gases under degenerate conditions.  One of the remarkable features of such gases is that interactions between atoms exhibit molecular resonances, whose energies can be tuned by, e.g., varying the external magnetic field.   It is therefore possible to realize experimentally gases in which correlations are strong, in the sense that the magnitude of the scattering length is comparable to or greater than the interparticle spacing.  At the 1999 conference on Recent Advances in Many-Body Theories, George Bertsch threw down the challenge of determining the properties of a Fermi system in which the magnitude of the scattering length is very much greater than the interparticle spacing \cite{bertsch}.  Under such conditions, the scattering length becomes an irrelevant parameter and the only length scale in the problem is the interparticle spacing, $r_s$.  As a consequence, the only relevant quantum-mechanical energy  scale for low-energy phenomena is $\hbar^2/2mr_s^2$, which for Fermi systems is equivalent to the Fermi energy.   Thus, in this regime, the energy per particle and the pairing gap are given by universal numbers times  $E_F$.   Specifically, the pairing gap is calculated to be $0.5 E_F$ \cite{unitarygap}.   

Does low-density neutron matter provide a realization of this universal behavior to be expected for a strongly-interacting system with short-range interactions?  At first sight, one might expect so: the density needed to achieve the condition $k_F |a|\gtrsim 1$ is not a problem, since this corresponds to densities of order $10^{-5}$ fm$^{-3}$.  However, one can see, e.g., from the behavior of the gap in the BCS approximation for cold atoms as a function of $k_F |a|$ shown in Fig.\ \ref{fig:bcscomp}, that the asymptotic behavior of the gap is achieved only for  $k_F |a|\gtrsim 10$.  However, the BCS gap for neutrons does not saturate, but begins to fall as $k_F |a|$ increases.  The reason is that, for Fermi wave numbers of order $10/|a| $, scattering of neutrons is not well approximated by the expression in which only the scattering length enters, and it is necessary to take into account momentum-dependent terms.  In the effective-range expansion, the $^1$S$_0$ phase shift is related to the scattering length and the effective range, $r_e$, by the expression
\be
\cot \delta_0=-\frac1a+\frac12k^2 r_e + \ldots.
\ee
Since the effective range for neutrons is $2.7$ fm, one sees that for $k_F\simeq 10/|a|$ the momentum-dependent terms must be included.  To express this result in other terms, if the leading term  ($-1/a$) in the effective-range expansion were sufficient, to achieve $k_F >10 /|a| $ would imply a phase shift of $\arctan 10\approx 84\degree$, while for neutron--neutron scattering the maximum phase shift
is only $65 \degree$ (see Fig.\ \ref{fig:phaseshifts}).\footnote{For neutrons, the effective-range approximation is good only for a limited range of momenta.  If one retains only the first two terms in the expansion, the phase shift is always positive for a negative scattering length and a positive effective range (the situation for neutrons) while the observed phase shift (Fig.\ \ref{fig:phaseshifts}) changes sign. Thus the vanishing of the gap at higher densities cannot be understood in terms of the effective-range expression for the phase shift.}  For atomic systems, large values of the scattering length may be achieved by exploiting molecular resonances, so-called Fano--Feshbach resonances, near zero energy in the two-atom system. Such resonances are referred to as broad if the two-body scattering amplitude is well approximated by its zero energy value, and as narrow when it is not \cite{PethickSmith}. For low momenta the behavior of the neutron--neutron  $^1$S$_0$ phase shift  bears a resemblance to that for atomic systems with a narrow resonance (however, for a narrow resonance the effective range is negative).

The overall picture of neutron matter that  emerges is that at low densities, $n\ll 1/|a|^3$, the effects of interactions may be calculated from a low-density expansion.  With increasing density, correlations initially become more important but for densities of order $1/(|a|r_e)^{3/2}$, the interaction between particles with momenta of order $k_F$ becomes weaker, and the pairing gap becomes smaller.  This has been exploited to calculate the energy of neutron matter at densities $n\sim n_s/10$ in a systematic expansion based on the scattering length and effective range \cite{SchwenkPethick}.  While at low densities the interaction may be approximated by a simple S-wave one, at higher densities the spin-dependent terms  and three-body contributions play an increasingly important role. 

\section{Microscopic calculations of pairing gaps}

In this section, we describe microscopic calculations of pairing gaps
in neutron matter.\footnote{Since our aim is to provide a critical review of state-of-the-art calculations, we
do not discuss the more historical results, for which we refer to
Ref.~\cite{page_this volume}.}  The simplest approach is to use the BCS approximation, in
which the pairing interaction is taken to be the free-space nucleon--nucleon interaction
and it is assumed that the single-particle energies are those of free
particles. Although this approximation does not reproduce the correct
weak-coupling result because, as discussed in Sec.~\ref{sec:inducedinteractions},  it does not include induced interactions, 
it does provide a useful benchmark, since
it probes the dependence on the nucleon--nucleon interaction used. Guided by the
insights from low densities and the comparison with ultracold atomic
gases, we shall discuss our understanding of the $^1$S$_0$ pairing gap
in neutron matter at low densities, focusing on quantum Monte Carlo
calculations that provide a systematic approach at strong coupling. At
higher densities the situation is less clear, and we shall provide a
critical review of calculations of the $^1$S$_0$ pairing gap in neutron
matter. Finally, we shall comment on the challenges involved in
microscopic calculations of $^3$P$_2$ neutron pairing and of proton
pairing in neutron stars.

\subsection{BCS approximation}
\label{sec:BCSapproximation}
There are a number of two-body interactions that fit low-energy nucleon--nucleon
scattering data and there is very good agreement between the gaps
calculated with these interactions although their short-distance
behaviors differ. The reason for the good agreement is that, for
calculating gaps, the quantity that matters is the scattering
amplitude at energies of order the Fermi energy, and this is strongly
constrained by nucleon--nucleon scattering data for nucleon momenta  in the center-of-mass frame $k \lesssim 2 \, {\rm
  fm}^{-1}$ \cite{Bogner:2003wn}.\footnote{For simplicity, we shall frequently adopt the common practice of working in units in which $\hbar=1$, in which case ``momentum'' and ``wave number" become synonymous.}  For higher momenta, there is considerable model
dependence, also because inelastic channels start to open up in nucleon--nucleon
scattering, e.g., pion production for $k > 1.7 \, {\rm fm}^{-1}$.

\begin{figure}[t]
\begin{center}
\includegraphics[clip=,width=0.9\textwidth]{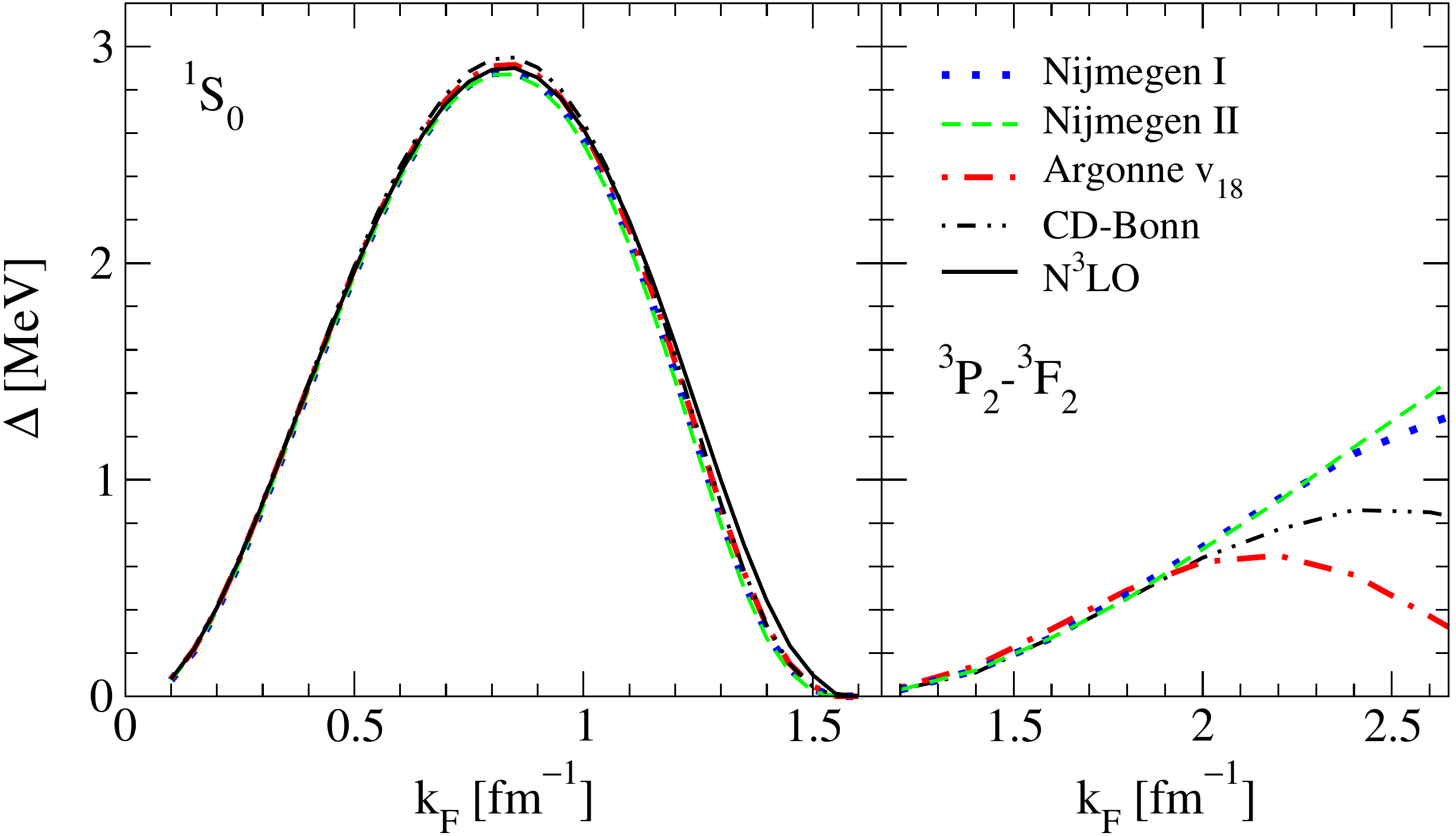}
\end{center}
\caption{The $^1$S$_0$ (left panel) and $^3$P$_2-^3$F$_2$
(right panel) pairing gaps $\Delta$ at the Fermi surface
as a function of Fermi wave number $k_F$ in neutron matter calculated in the BCS approximation for a number of  charge-dependent nucleon--nucleon interactions that have been fitted to nucleon--nucleon scattering data.  The potentials are specified according to the legend in the right panel.
For details see Refs.~\cite{Hebeler:2006kz} (left panel) and    \cite{Baldo:1998ca} (right panel).
\label{BCSgaps}}
\end{figure}

Figure~\ref{BCSgaps} shows the $^1$S$_0$ and $^3$P$_2-^3$F$_2$ pairing
gaps in neutron matter, obtained by solving the BCS gap equation with
a free-particle spectrum for the normal state. At low densities (in the crust of neutron stars),
neutrons form a $^1$S$_0$ superfluid. At higher densities, the S-wave
interaction is repulsive and neutrons pair in the $^3$P$_2$ channel
(with a small coupling to $^3$F$_2$ due to the tensor
force). Figure~\ref{BCSgaps} demonstrates that in the BCS approximation the $^1$S$_0$ gap
is essentially independent of the nuclear interaction used~\cite{Hebeler:2006kz}. This
includes a very weak cutoff dependence for low-momentum interactions
$\vlowk$. The inclusion of the leading three-nucleon forces in chiral effective field theory gives a reduction of
the $^1$S$_0$ BCS gap for Fermi wave numbers $k_F > 0.6 \, {\rm
  fm}^{-1}$~\cite{Hebeler:2009iv}. This reduction becomes significant
for densities where the gap is decreasing and agrees qualitatively
with results based on three-nucleon potential models (see, e.g.,
Ref.~\cite{Zuo:2004mc}). At low densities ($k_F \lesssim 0.6 \, {\rm
  fm}^{-1}$), $^1$S$_0$ pairing can therefore be calculated using only two-body
interactions. At higher
momenta, model potentials lead to different predictions for nucleon--nucleon scattering and this shows up prominently in Fig.~\ref{BCSgaps} in the
$^3$P$_2-^3$F$_2$ gaps for Fermi wave numbers $k_F > 2 \, {\rm
  fm}^{-1}$~\cite{Baldo:1998ca}.

\subsection{Low densities}

At low densities, great progress has been made in calculating pairing
gaps including medium effects using quantum Monte Carlo methods, a family of
techniques used in condensed-matter physics, materials science,
quantum chemistry, and nuclear physics to solve the quantum many-body
problem using a stochastic approach (for a review see
Ref.~\cite{foulkes}). This is not the place for a detailed account of
these methods but we hope to be able to give the reader a sense of the
basic ideas behind them and of the ingredients needed to obtain
reliable results.  The methods are not restricted to weak potentials,
for which perturbative methods are applicable, and state-of-the-art
implementations of them lead to results for the energy that are accurate to within
1\% of the value obtained from experiment or by exact
diagonalization.  For simplicity we shall limit ourselves to the case
of zero temperature.

In the simplest of these methods, Variational Monte Carlo (VMC), one
constructs for the $N$-particle system a trial wave function $| \psi_V
\rangle$, which is a function of the $3N$ coordinates.  The Hamiltonian $H$ of the system contains contributions
from the kinetic energy, and from two-body interactions (and possibly
also three- and higher-body interactions), and one evaluates the
integrals that occur in the expectation value of the Hamiltonian by
stochastic sampling in the multi-dimensional space.  From the
Rayleigh-Ritz principle it follows that
\begin{equation}
E_{\text{VMC}} \equiv \frac{ \langle \psi_V | H | \psi_V \rangle}
{ \langle \psi_V | \psi_V \rangle} \geq E_0\,,
\end{equation}
where $E_0$ is the ground-state energy.  By varying parameters in the
trial wave function as well as its functional form, one can find the
wave function that gives the lowest energy.

To see how one may further improve the wave function, consider the evolution of a state of the system in time.  This is given by the equation
\be
i\hbar \frac{\partial | \psi \rangle}{\partial t}=H| \psi \rangle\,.
\ee
One may expand the wave function in terms of energy eigenstates $| \psi_n \rangle$  of the Hamiltonian with eigenvalue $E_n$:
\be
| \psi(t) \rangle=
\sum_n c_ne^{-iE_nt/\hbar}| \psi_n \rangle,
\ee
where the $c_n$ are coefficients independent of time.  
If one introduces the imaginary time variable $\tau=it$, one sees that with increasing $\tau$, the component having the largest amplitude for large $\tau$ is that for the ground state.  Thus, evolution of the wave function  in imaginary time, which satisfies the equation 
\be
 -\hbar\frac{\partial | \psi \rangle}{\partial \tau} = H |\psi\rangle,
 \label{diff}
\ee
systematically ``purifies'' the state by preferentially removing
excited state contributions.  This provides the basis for the
Diffusion Monte Carlo (DMC) method, whose name reflects the fact
that in coordinate space the kinetic energy operator is proportional
to a sum of terms of the form ${\nabla_i^2}$, where $i$ is the
particle label, and therefore Eq.\ (\ref{diff}) resembles a diffusion
equation in a $3N$ dimensional space.\footnote{We caution the reader
  that there is no generally agreed upon nomenclature for the various
  Monte Carlo methods.  To keep the discussion simple we shall not
  distinguish between the DMC method and the closely related Green's
  Function Monte Carlo (GFMC) one.}  In practice, the starting point
for the DMC method is frequently the wave function obtained from a VMC
calculation.

Quantum Monte Carlo simulations for many-boson systems are, in
principle, exact: an input wave function is employed but its only role
is to reduce the statistical variance of the final result. Fermions
are different due to the  ``fermion sign problem''.
This arises because the ground state of the many-fermion problem
corresponds to an excited state of the many-boson problem. When
propagated in imaginary time, an initially antisymmetric wave function
will, due to the statistical sampling, acquire
components that are symmetric in the particle coordinates, and with
time these grow and dominate the wave function: an initially fermionic
wave function will thus evolve to the bosonic ground state.  One
method to circumvent this difficulty is to use a ``fixed-node''
approximation (for real wave functions) or a ``constrained path'' one
(for complex wave functions), in which the stochastic evolution is
artificially constrained~\cite{ceperley}.  For simplicity, we describe
the method for real wave functions.  Because of the antisymmetry of
the spatial wave function, it has positive and negative regions, which
are separated by nodal surfaces, which divide the multi-dimensional
configuration space into a number of domains in which the wave
function does not change sign.  Within these domains, the evolution of
the wave function corresponds to that of a many-boson problem.  In the
fixed-node approximation, one solves the evolution within each domain
separately, keeping the positions of the nodal surfaces fixed.  Unlike
in the variational Monte Carlo approach, the wave function within each
domain is not constrained to have a particular functional form.  In the fixed-node approximation one needs to specify the nodal structure in some way and different choices lead to different ground-state energies.

Due to the computational demands, quantum Monte Carlo calculations can
only be carried out for systems composed of (at most) a few hundred
particles. To obtain results for
the thermodynamic limit ($N \rightarrow \infty$, $\Omega \rightarrow
\infty$, with $N/\Omega \rightarrow \text{constant}$, $\Omega$ being the volume of the system) the dependence of
results on the particle number $N$ must be carefully
studied. Generally speaking, if the range of the interaction is small
compared with the particle spacing, relatively few particles are
needed to simulate the infinite system: e.g., for cold alkali
atoms, where the particle spacing is $\sim 10^{-4}$ cm while the range
of the interaction is of order $10^{-6}$ cm, there is essentially no
variation when the particle number is increased above 40 (at zero
temperature). The situation is different in the case of nuclear physics,
since the range of the interaction ($\sim 1$ fm) is comparable
to the interparticle spacing.

A second complicating feature of nuclear physics is the spin and
isospin dependence of the interaction. As a consequence of the rapid 
increase of the
number of spin states with particle number, it is at
present possible to study systems with 14 neutrons (or 12 nucleons if both
neutrons and protons are present) \cite{Pieper:2008}.

A promising method, referred to as Auxiliary Field Diffusion Monte
Carlo (AFDMC)~\cite{AFDMC} extends the stochastic evolution in
coordinate space in the DMC method to spin-isospin space.  This
exploits the Hubbard--Stratonovich identity \cite{HS}, which expresses
a two-body operator as a sum of one-body operators interacting with
random fields (the auxiliary fields in the name for the method) and integration over these fields.  This  
integration is performed by Monte Carlo techniques analogous to those used to study the wave
function in coordinate space.  This method has the advantage that it
can be used for larger systems but it suffers from the disadvantage
that it does not give an upper bound on the energy.

In nuclear physics, a measure of the pairing gap is the systematic staggering of
ground-state energies between nuclei with even numbers of neutrons (or
protons) and those with odd numbers.  Explicitly for the case of
neutrons, in systems with even neutron number, $N$, all neutrons are
paired, while for odd $N$ one nucleon is not paired.  The simplest way
to express this pairing gap is in terms of the second difference
\begin{equation}
\Delta = (-1)^{N+1}\left\{
E(N) - \frac{1}{2} \left [ E(N+1)+E(N-1) \right ]\right\}, 
\label{gapNP}
\end{equation}
where $E(N)$ is the ground-state energy of a system with $N$ neutrons.   Equation (\ref{gapNP}) may also be written in terms of the neutron separation energy, 
\be
S(N)=E(N)-E(N-1),
\ee
as
\be
\Delta = \frac{(-1)^N}{2}[
S(N+1) - S(N)].
\ee
For large systems, this is equivalent to the definition in terms of the gap calculated within the BCS theory and extensions of it.  In that approach, the minimum energy necessary to add an excitation to the system without changing the average particle number is $\Delta$.  Addition of a neutron to the ground state of a system with an even number of neutrons requires an energy $\mu+\Delta$, where $\mu$ is the chemical potential while to remove a particle requires an energy $-\mu+\Delta$.  In both cases, one neutron is unpaired, so an extra energy $\Delta$ is required over and above the energy $\pm\mu$ required to add or remove a particle in the absence of pairing.  One thus sees that the definition (\ref{gapNP}) is consistent with this.
For applications to finite nuclei it is common to use higher-order
difference expressions (see, e.g., \cite[p.~17]{BrinkBroglia}) in
order better to remove $N$-dependent contributions to the energy not
due to pairing. However, in the case of unbound systems such as 
neutron matter, for which the dependence of the energy on particle
number is smoother, this extra complication is unnecessary.
Calculation of the pairing gap demands high accuracy for the
calculations of the ground-state energies, since the pairing gap is
small compared with the total energy of the system, $\sim NE_F$.  Thus, to
resolve the pairing gap, statistical errors must be reduced well below
a level of $\sim \Delta/(N E_F)$.  For low density neutron matter
$\Delta/E_F$ can reach values as high as $~ 0.4$ and gaps can be
calculated reliably, while in terrestrial superconductors $\Delta/E_F$
is much smaller and gaps cannot be extracted from numerical
calculations of ground-state energies.

The simplest choice of a variational wave function, $|\psi_V \rangle$,
for a normal gas is a Slater determinant, $|\psi_S \rangle$, of
single-particle orbitals chosen according to the problem at hand:
for neutron matter, they are plane waves. One typically also includes
Jastrow correlation factors of the form $\prod_{i<j} f(r_{ij})$,
possibly multiplied by an operator relevant to the interaction (in
nuclear physics this includes spin-dependent central, spin-orbit, and tensor terms).  Both the DMC and the
AFDMC methods work with Jastrow factors that are only central: for
neutron matter at low densities, where the interaction is mainly
S-wave, this is expected to be a reasonable approximation. We noted
earlier that the nodal structure of the variational wave function influences the final result. This also means that for functions
$f(r)$ that are nodeless, the Jastrow term is relatively unimportant, except for
controlling the statistical error.

For superfluid systems, pairing correlations strongly affect the nodal structure of the wave function and must be incorporated in the trial wave function from the start. The simplest choice is the BCS wave function (\ref{eq:bcswave}), which may be written as 
\begin{equation}
| \psi_{BCS} \rangle \propto  \exp\left(\sum_{\bf k} (v_{{\bf k}}/u_{{\bf k}}) c^{\dagger}_{{\bf k} \uparrow}
c^{\dagger}_{{-\bf k} \downarrow}\right)|0\rangle~,
\label{eq:bcswaveEXP}
\end{equation}
since a fermion creation or annihilation operator raised to a higher  power than unity vanishes.  Here $|0\rangle$ is the vacuum state.
Thus, the  component of this wave function with a definite (even) number of particles $N$ may be written as 
\begin{equation}
| \psi_{BCS,N} \rangle \propto  \left(\sum_{\bf k}\alpha_{\kk} c^{\dagger}_{{\bf k} \uparrow}
c^{\dagger}_{{-\bf k} \downarrow}\right)^{N/2}|0\rangle~,
\label{eq:bcswaveEXP_N}
\end{equation}
where $\alpha_{\kk}=v_{{\bf k}}/u_{{\bf k}}$.  The wave function thus
corresponds to the antisymmetrized product of $N/2$ pairs of particles
in the state specified in momentum space by the coefficients
$\alpha_{\kk}$.  This wave function may be used as a trial function
for general $\alpha_{\kk}$, and the optimal form will generally be
different from the BCS mean-field result.  In coordinate space, the
pair wave function is the Fourier transform of $\alpha_{\kk}$.

\begin{figure}
\begin{center}
\includegraphics*[width=0.8\columnwidth]{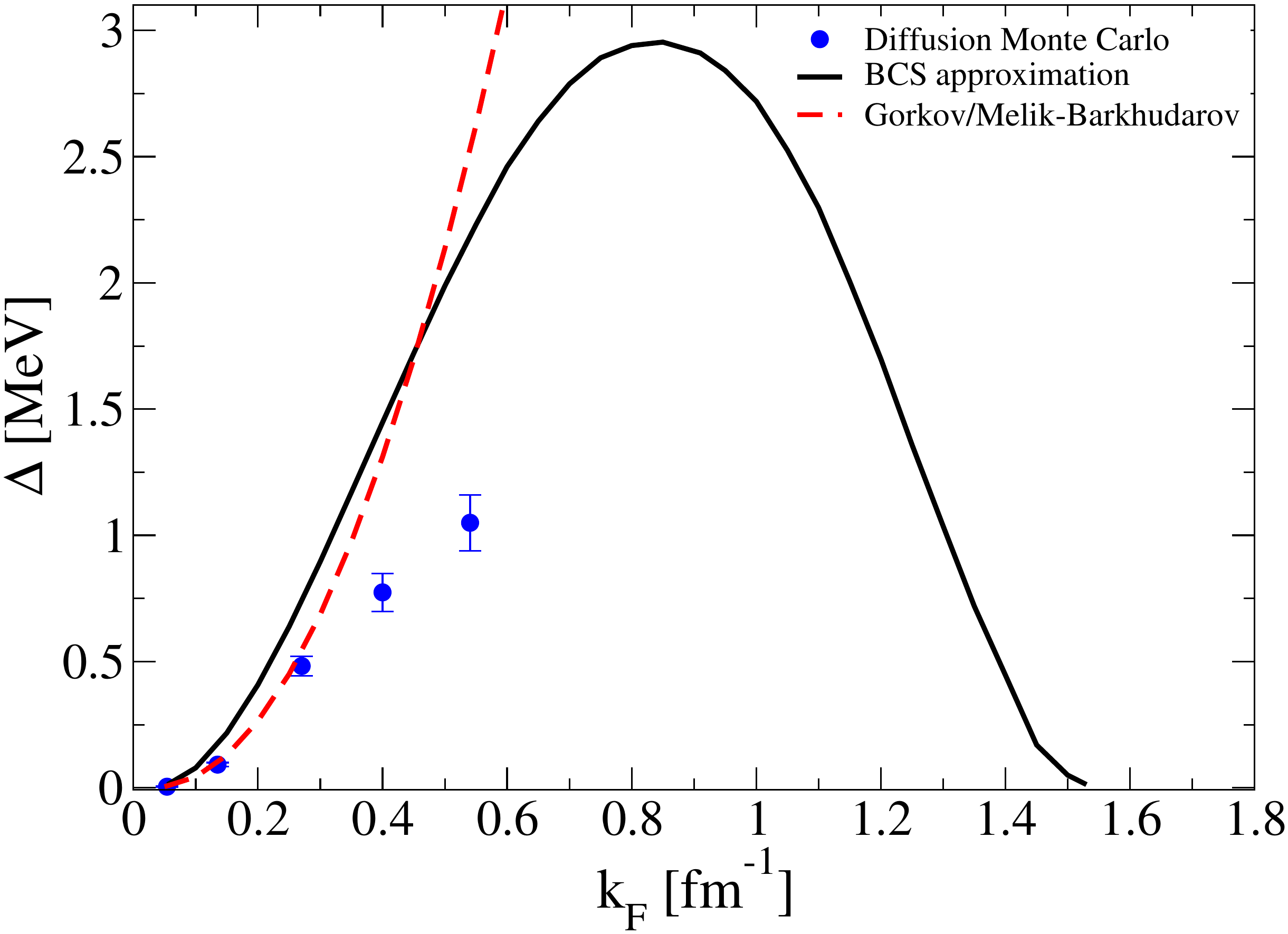}
\end{center}
\caption{The $^1$S$_0$ pairing gap $\Delta$ as a function of Fermi wave number $k_F$
in neutron matter. Results are shown for the BCS approximation
(see Fig.~\ref{BCSgaps}), for the exact result in the low-density limit (Gor'kov/Melik-Barkhudarov)~\cite{Gor'kov}, which includes induced interactions, and
for Diffusion Monte Carlo calculations of neutron 
matter~\cite{Gezerlis2008}.\label{GezerlisGaps1}}
\end{figure}

We turn now to explicit calculations of gaps in neutron matter.  One
general point is that, due to the standard approaches to handling the
fermion-sign problem (as detailed above), the nodal structure of the
input wave function can really influence the final answer. For
example, in the unitary Fermi gas $|\psi_S \rangle$ gives a
ground-state energy that is 20\% or more larger than that from $| \psi_{BCS,N}
\rangle$ \cite{Carlson}. Even when a BCS-like wave function is used, there may be
considerable dependence on the specific values chosen for
$\alpha_{\kk}$. This has, indeed, been the case for low-density
neutron matter: since the interaction at such densities is dominated
by its S-wave component, it is possible to apply both DMC and AFDMC,
but the results obtained from the two approaches differed, due to the difference
in input wave functions. Specifically, the DMC calculations in
Ref.~\cite{Gezerlis2008} used a variationally optimized wave function,
leading to the results shown in Fig.~\ref{GezerlisGaps1}.  An important
check on numerical calculations is provided by the analytic results of
Gor'kov and Melik-Barkhudarov \cite{Gor'kov} described in
Sec.\ \ref{sec:dilute} and also shown in the figure. The fact that
such DMC calculations agree with the expected analytic behavior at the
lowest densities, namely a suppression with respect to the BCS
approximation value by a factor $(4 e)^{-1/3}\approx 0.45$, increases one's
confidence that they behave properly at intermediate densities.
Another set of quantum Monte Carlo calculations used AFDMC along with
$\alpha_{\kk}$ calculated within the Correlated Basis Function (CBF)
approach \cite{Fabrocini2005}.\footnote{The calculations included in
  addition a three-body interaction described by the Urbana IX 
  potential but at low densities this should play no role, as discussed in Sec.\  \ref{sec:BCSapproximation}} This CBF
wave function is ill-behaved since at very low densities it does not reproduce the results of Gor'kov and Melik-Barkhudarov. 
Since these CBF wave functions were used as input and define the nodal structure in the AFDMC computations, the AFDMC results also do not reproduce the correct low-density limit.

The present status of calculations of gaps in neutron matter is that
the situation at densities $\lesssim n_s/10$ is under good control,
thanks to the analytical results in the low-density limit and quantum Monte
Carlo methods: the physical reason for the suppression of the gap
compared with the BCS approximation is the repulsive interaction
induced by exchange of spin fluctuations.  At higher densities there
are larger uncertainties because additional terms in the
neutron--neutron interaction become increasingly important and the increased density makes  the extraction of gaps from energy differences more challenging.

 In addition,
three-neutron interactions begin to play a role but for neutron matter
their effects are suppressed because configurations in which three
neutrons are close together are unlikely, since at least two of the
neutrons must be in the same spin state.  

\subsection{Higher densities}

\begin{figure}
\begin{center}
\includegraphics*[width=0.8\columnwidth]{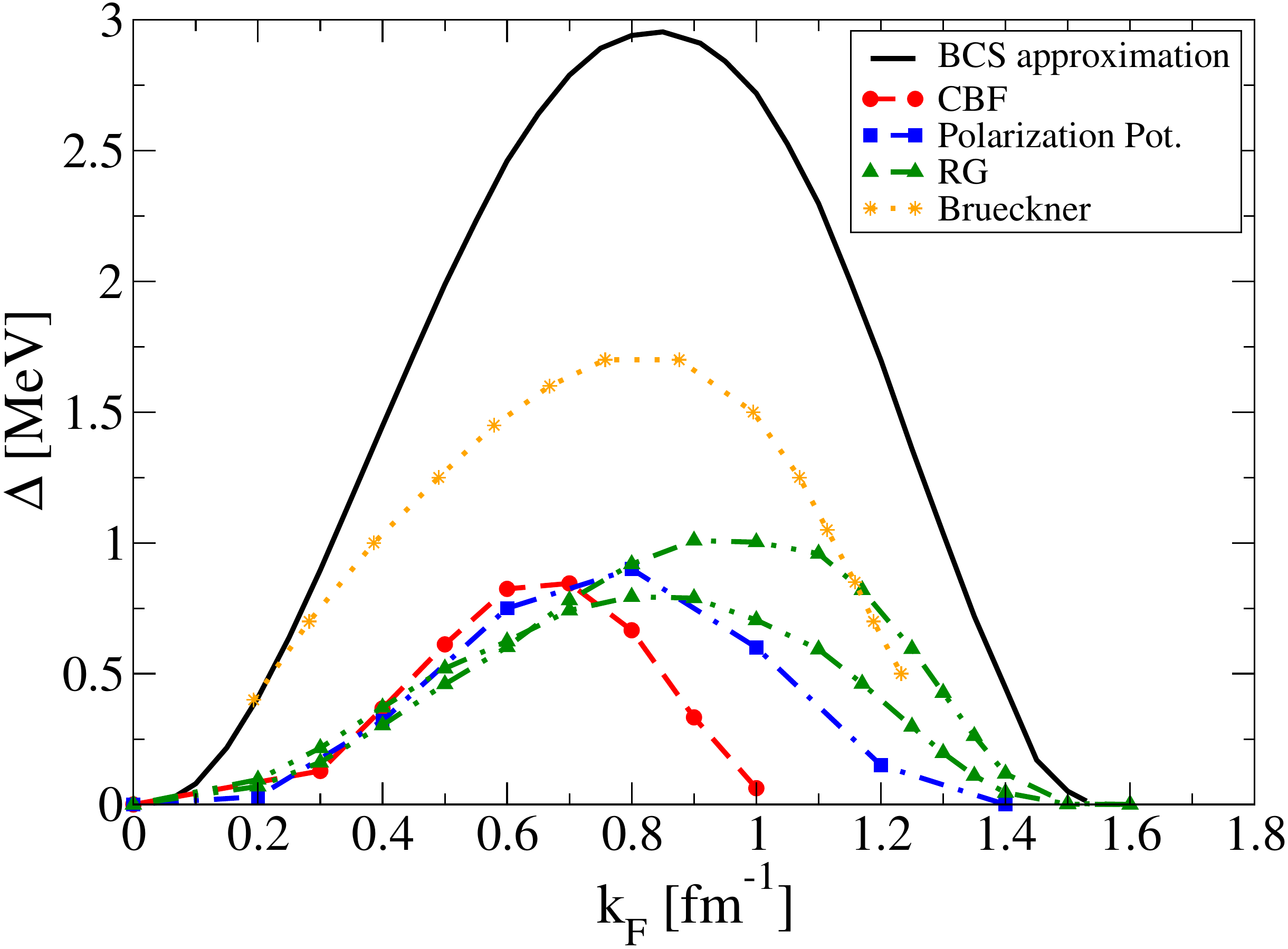}
\end{center}
\caption{The $^1$S$_0$ pairing
gap $\Delta$ at higher densities as a function of Fermi wave number $k_F$. Results are shown for the BCS approximation
(see Fig.~\ref{BCSgaps}), for the method of Correlated Basis Functions
(CBF)~\cite{clark}, for the polarization potential method, in which induced interactions are calculated in terms of pseudopotentials (Polarization Pot.)~\cite{Wambach:1993}, for a calculation in which induced interactions in the particle--hole channels are calculated from a
renormalization group (RG) approach~\cite{Schwenk:2002fq}, and for calculations based on Brueckner
theory~\cite{CaoLombardoSchuck}.\label{GezerlisGaps2}}
\end{figure}

Figure~\ref{GezerlisGaps2} demonstrates that understanding many-body
effects beyond the BCS approximation constitutes an important open
problem at higher densities. All calculations shown in
Fig.~\ref{GezerlisGaps2} are based on nucleon--nucleon interactions only, so the
differences are due to truncations in the many-body calculations.

As discussed in Sec.\ \ref{sec:inducedinteractions}, induced interactions due to screening and
vertex corrections (creation of particle--hole pairs in intermediate states) are crucial for a quantitative understanding of pairing gaps. They lead to a
reduction of the $^1$S$_0$ gap that is significant even in the
limit of low densities,  $k_F |a|\to 0$ ~\cite{Gor'kov}:
\begin{eqnarray}
\Delta &=& \frac{8}{e^2} \, E_F \exp \biggr\{
\, {\rm const.} \, \biggr( \, \begin{minipage}{4.4cm}
\includegraphics[scale=0.5,clip=]{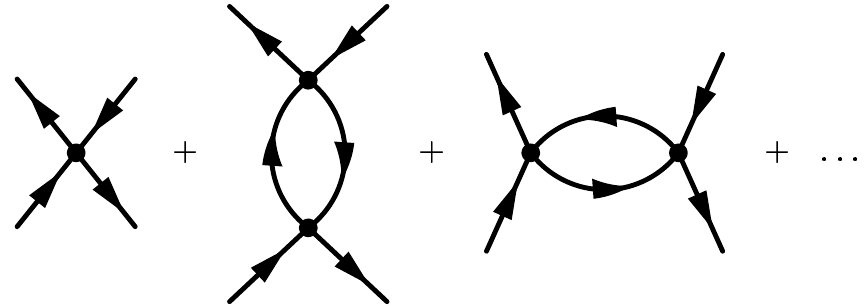}
\end{minipage}\biggr)^{-1} \biggr\} \nonumber \\
&=& \frac{8}{e^2} \, E_F \exp\biggl(\frac{\pi}{2 k_F a} 
+ \ln[(4e)^{-1/3}] + {\mathcal O}(k_F a)\biggr) \,.
\end{eqnarray}
Because the exponent depends  nonperturbatively on the pairing interaction, $\sim 1/g$,  second-order induced interactions in the
particle-hole channels lead to an $\mathcal{O}(1)$ correction to the
pairing gap, $1/(g + C_2\, g^2) \simeq 1/g - C_2$ in weak coupling,
which leads to a reduction by a factor $(4e)^{-1/3} \approx 0.45$ in very low density neutron matter.

Following Shankar~\cite{Shankar:1993pf}, Ref.~\cite{Schwenk:2002fq}
developed a nonperturbative renormalization group (RG) approach for
neutron matter, where induced interactions in the particle-hole
channels are generated by integrating out modes away from the Fermi
surface. Starting from a nucleon--nucleon interaction, the solution to the RG
equations in the particle-hole channels includes contributions from
successive particle-hole momentum shells. The RG builds up many-body
effects in a way similar to that in which the two-body parquet equations do, and efficientlyi
includes induced interactions to low-lying states in the vicinity of
the Fermi surface beyond a perturbative calculation. For
momentum-independent interaction vertices, one can argue that the
resummation of the particle-hole channel agrees with the truncated
weak-coupling expansion: schematically after resumming the
particle-hole channels: $1/(g + C_2\, g^2 + \ldots) \to (1 - C_2 \,g)/g =
1/g - C_2$. For a discussion in the context of the BCS-BEC crossover,
see Refs.~\cite{GubbelsStoof,Diehl}.

The RG results~\cite{Schwenk:2002fq} for the $^1$S$_0$ pairing gap based on a low-momentum interaction
$\vlowk$ are
shown in Fig.~\ref{GezerlisGaps2}. Induced interactions reduce the maximal gap by a factor $\sim 3$ 
to $\Delta \approx 0.8 \mev$. The two RG results shown in
Fig.~\ref{GezerlisGaps2} provide a measure of the uncertainty due to
an approximate treatment of the neutron self-energy. At low densities, the RG approach
is consistent with the result $\Delta/\Delta_{\rm BCS} \approx 0.45$ in the dilute limit,
see also Ref.~\cite{Schwenk:2006tz}. Note that the RG results give
smaller gaps than the DMC results in Fig.~\ref{GezerlisGaps2}. This is
because the RG results are obtained from a weak-coupling BCS formula
with pairing interaction from the particle-hole RG, and the
weak-coupling BCS gaps are in reasonable agreement with the BCS
approximation except at lower densities (see Fig.~8 in
Ref.~\cite{Schwenk:2002fq}).

\begin{figure}[ht]
\begin{center}
\includegraphics[width=0.55\textwidth,clip=]{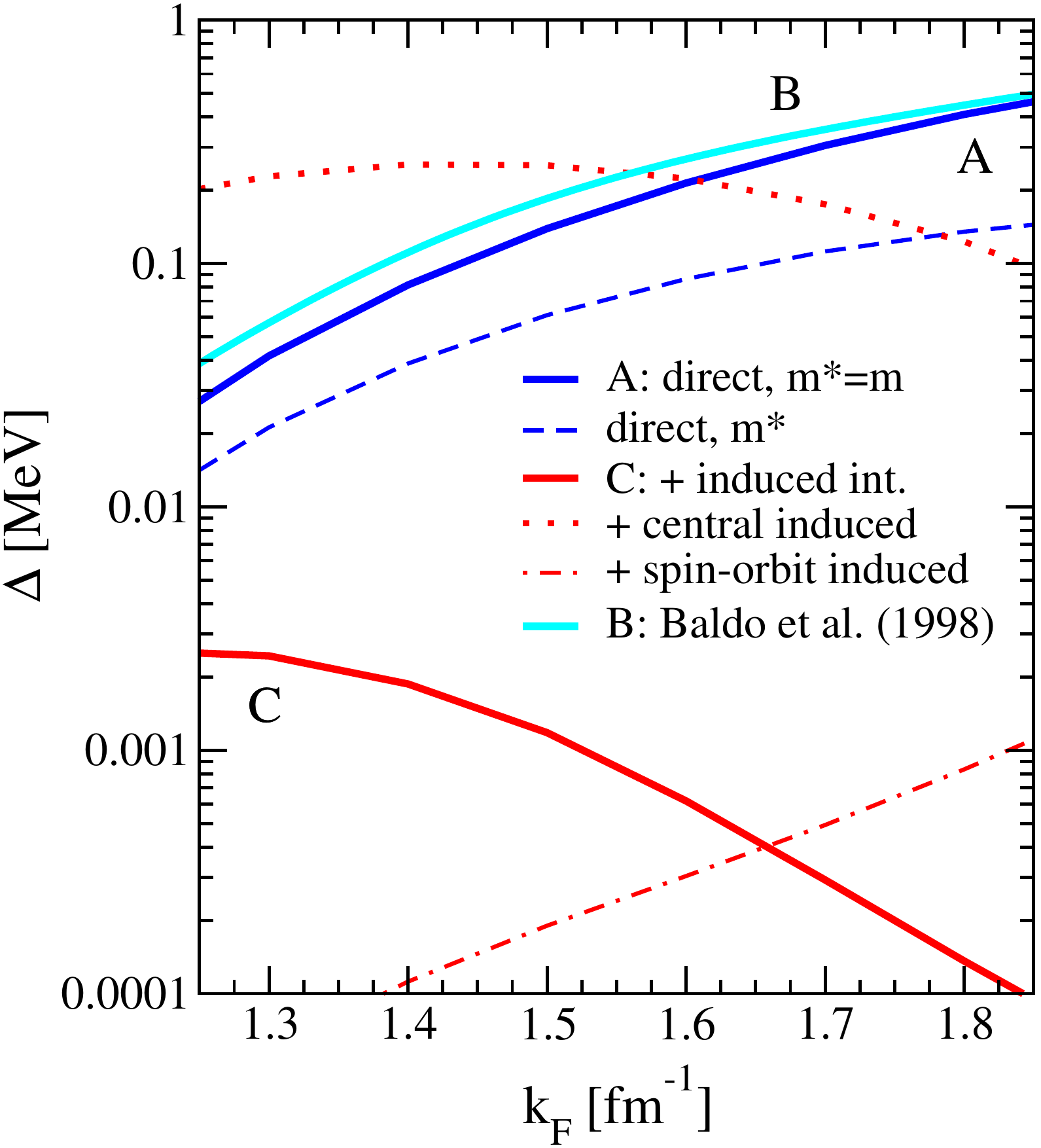}
\end{center}
\caption{Angle-averaged $^3$P$_2$ pairing gap $\Delta$
versus Fermi wave number $k_F$ in neutron matter. The calculations were made in the spirit of the BCS schematic model and for weak-coupling, and the gap is given by 
$\Delta=2E_F\exp(1/N(E_F) V)$, where $V$ is the pairing matrix element in the $^3$P$_2$ channel at the Fermi surface.  The results are shown for the direct interaction only, both without and with corrections to the effective mass of the particles, and with the inclusion of second-order induced interactions in 
the pairing interaction. We also show the 
modification of the gap when only induced central or only induced
spin-orbit effects are taken into account. For comparison, we give the
results of Baldo {\it et al.}~\cite{Baldo:1998ca}, which are obtained
by solving the BCS gap equation in the coupled $^3$P$_2$--$^3$F$_2$
channel for different free-space interactions (see also Fig. \ref{BCSgaps}).  For details 
see Ref.~\cite{SchwenkFriman}.\label{3p2gaps}}
\end{figure}

Figure~\ref{GezerlisGaps2} shows that gaps calculated with Correlated Basis
Function theory~\cite{clark} and with induced interactions based
on pseudopotentials~\cite{Wambach:1993} are in
reasonable agreement with the RG results at lower densities, but there
are considerable spreads in the density at which the gap is maximal and in the density at which it vanishes. In addition, in
Fig.~\ref{GezerlisGaps2} we show results for the $^1$S$_0$ pairing
gap in Brueckner theory~\cite{CaoLombardoSchuck}, but they 
disagree with the known analytical result at low densities, where they predict a gap greater than $\Delta_{\rm BCS}$.

\subsection{$^{\bf 3}$P$_{\bf 2}$ pairing of neutrons}

Noncentral spin-orbit and tensor interactions are crucial for
$^3$P$_2$ superfluidity. In particular, without an attractive
spin-orbit interaction, neutrons would form a $^3$P$_0$ superfluid, in which the spin and orbital angular momenta are anti-aligned, 
rather than the $^3$P$_2$ state, in which they are aligned.

Gaps calculated in the BCS approximation are shown in Fig.~\ref{BCSgaps} and
calculations of $^3$P$_2$ pairing including many-body
effects are given in Refs.~\cite{Zhou,Khodel:2004nt,Dong:2013sqa,PethickRavenhall}. 
The $^3$P$_2$ gap is more sensitive to the pairing interaction than the S-wave gap, 
because the $^3$P$_2$ gap is very small compared to the Fermi energy.
To date, there is
only one (perturbative) calculation of non-central induced
interactions~\cite{SchwenkFriman}. This showed that, due to the
coupling of the tensor and spin-orbit force to the strong spin-spin
interaction ($G_0$), the tensor component of the quasiparticle
interaction and the (P-wave) spin-orbit interaction are significantly
reduced in neutron matter. As a result, $^3$P$_2$ gaps below $10 \,
{\rm keV}$ are possible in the interior of neutron stars, see
Fig.~\ref{3p2gaps}~\cite{SchwenkFriman}. For $^3$P$_2$ pairing it is
crucial to include non-central induced interactions, because P-wave
pairing would be enhanced if only central induced interactions were
included~\cite{PethickRavenhall}. Note that these calculations do not include three-nucleon forces, which were found to increase the $^3$P$_2$ pairing gaps~\cite{Zhou}.
Understanding three-body forces is a frontier area in nuclear physics (for a recent review see Ref.~\cite{Hammer:2012id}).

Typical interior temperatures of isolated neutron stars are on the
order of $10^8 \, {\rm K} \approx 10 \, {\rm keV}$ and
therefore it is possible to constrain the $^3$P$_2$ pairing gaps
phenomenologically through neutron star cooling
simulations~\cite{page_this volume,YakovlevPethick}. The small
$^3$P$_2$ pairing gaps in Fig.~\ref{3p2gaps} would imply that core
neutrons are only superfluid at late times $(t \gtrsim 10^5 \, {\rm yrs})$.

\subsection{Proton pairing}

In neutron stars, matter is in beta equilibrium, with equal rates for neutrons undergoing
beta decay and protons capturing electrons. As a result, matter in the interior of
neutron stars at densities comparable to that of nuclear matter consists of $\sim 95\%$ neutrons and $\sim
5\%$ protons and electrons. Therefore, proton Fermi wave numbers are
considerably lower than neutron ones, $k_F^p = k_F^n [x/(1-x)]^{1/3}$, where $x=n_p/n$
is the proton fraction.  If only the free-space interaction contributed to pairing, protons in neutron stars would form a
$^1$S$_0$ superfluid, unless proton densities high enough to favor  $^3$P$_2$ pairing are reached.

Calculating proton pairing gaps is a challenging problem due to the
coupling of protons to the denser neutron background. An important effect is that
the proton effective mass, and therefore the density of states that
enters the pairing strength in the exponent in the expression for the gap, is smaller than the
neutron effective mass in neutron-rich matter, $m_p^* < m_n^*$ (see
the general considerations in Fermi liquid theory in
Ref.~\cite{Sjoeberg}). This decreases the proton $^1$S$_0$ gap
significantly for a particular proton density, ${n}_p$, compared to the neutron gap for a neutron density equal to $n_p$~\cite{Zhou,BaldoSchulze}. In addition to effective-mass
effects, the coupling to the denser neutron background amplifies the
repulsive contributions to the pairing interaction from three-nucleon forces and decreases the proton $^1$S$_0$ gap further~\cite{Zhou,BaldoSchulze}.

 The situation regarding induced interactions is less clear. Reference \cite{WAP91}
found induced interactions to be repulsive
with a stronger reduction of the proton $^1$S$_0$ gap compared to the
neutron one, whereas Ref.~\cite{BaldoSchulze} found attractive induced interactions. A systematic study of proton pairing in
neutron-rich matter incorporating the discussed many-body effects
consistently remains an important outstanding problem.

\section{The inner crust}
\label{sec:innercrust}

From the calculations described above, one sees that pure neutron
matter is expected to be paired in the $^1$S$_0$ state at neutron
densities less than about $n_s/2$.  However, in neutron stars one does
not find bulk neutron matter at such densities, because this density
range corresponds to the inner crust, in which the neutrons permeate a
crystal lattice of neutron-rich nuclei.  The inner crust of neutron
stars, while making up only a small fraction of the mass of the star,
is important for a variety of reasons (see, e.g., Ref.~\cite{Haensel}).  One is that heat from the
interior of the star has to pass through this region on its way to the
surface, and therefore the transport properties of the matter
influence surface temperatures.  Another reason is that a number of
observed phenomena, including glitches in the rotation rate of pulsars
and quasiperiodic oscillations observed in giant flares emitted by
magnetized neutron stars, have been attributed to processes involving
a neutron superfluid in the inner crust. Thus, it is necessary to
understand the thermodynamic and transport properties of matter in the
crust.

\subsection{Static properties}
\label{sec:staticproperties}

Calculating the static properties is a formidable challenge for a
number of reasons.  As we have seen in earlier sections, even at the
relatively low densities of interest in neutron star crusts,
correlations are important.  This means that in a mean-field
calculation one must employ effective interactions rather than free-space
ones.  Typically, such interactions are fitted to observed properties of
nuclei and to the calculated properties of pure neutron
matter.  Particularly popular choices are of the
Skyrme form, which makes for computational simplicity, supplemented by
a pairing interaction.  In nuclear physics parlance, mean-field
calculations with effective interactions are referred to as
Hartree--Fock (HF) ones when pairing is neglected and as
Hartree--Fock--Bogoliubov (HFB) ones when it is included.  In the
HFB calculations, one solves self-consistently for the functions
$u_i(\rr)$ and $v_i(\rr)$ describing the particle and hole amplitudes
for excitations in the state $i$.  These two functions are the
analogues of the quantities $u_\kk$ and $v_\kk$ for infinite matter.

To date, no HFB calculations that take into account the crystalline
structure have been made and it is necessary to make further
assumptions.  A common one is to use what is referred to as the
Wigner--Seitz approximation, in which one replaces the unit cell in
coordinate space by a sphere of the same volume.\footnote{The name for
  this approximation stems from a similar approximation made in
  momentum space to simplify calculations of band structure
  \cite{WS}.}  To mimic the situation in a crystal, one applies
boundary conditions at the edge of the Wigner--Seitz cell that allow
the density there to remain nonzero, whereas for box boundary
conditions the neutron density there would vanish.  The present state
of the art is described in Ref.\ \cite{PastoreBaroniLosa} and a
comparison of results for a variety of effective interactions is given
in Ref.\ \cite{PastoreMargueron}.  A simplified way of calculating
pairing effects is to first perform a HF calculation for the normal
state and then to include pairing in the spirit of the BCS theory for
uniform matter, in which pairing occurs between particles in
time-reversed single-particle states which, e.g., for S-wave pairing means particles
with equal and opposite momentum and spin. We shall refer to this
approximation as ``HF+BCS''. More generally, pairing can occur between
particles in states that are not related by time reversal: in an atomic
nucleus pairing can occur between states with the same total
orbital angular momentum,  opposite projections of the orbital angular momentum and spin, but different radial quantum
numbers, so-called ``off-diagonal'' pairing.  The importance of these
off-diagonal terms is well illustrated by calculations of the specific
heat of matter in the inner crust \cite{PastoreThermal}. In the HF+BCS
approximation there is a single transition temperature to the paired
state.  However, in the inner crust, one has two rather different
sorts of matter (the nuclei and the interstitial neutrons), and if
coupling between the two components were small, one would expect there
to be two transition temperatures, one corresponding to the matter in
nuclei and the other corresponding to that for the outside neutrons.
The HFB results indeed exhibit two maxima in the specific heat,
corresponding to the transitions in the two sorts of matter.

\subsection{Hydrodynamic equations}
\label{sec:hydrodynamics}

To describe dynamical processes for matter in the inner crust is
generally complicated, but for long-wavelength and low-frequency
phenomena one can adopt a hydrodynamic approach.  Because of the
presence of a neutron superfluid in the lattice of nuclei in the inner
crust, the system has an extra degree of freedom compared with normal
matter.  Following Ref. \cite{cjpchamelreddy}, we now describe how one
may develop a two-fluid description in which the system is regarded as
being made up of a superfluid component and a normal component, as was
previously done for superfluid liquid $^4$He. While in the case of
liquid $^4$He the normal component corresponds to thermal excitations,
for matter in the inner crust, the normal component is made up of the
nuclei and electrons and so does not disappear at zero temperature.
The system thus has similarities to dilute solutions of $^3$He in
$^4$He, where the $^3$He atoms play the role of the normal component.
The state of the system locally is specified by the density of
neutrons, $n_n$, the density of protons, $n_p$, averaged over length
scales large compared with the spacing between nuclei, and the
velocity of nuclei (and of electrons which move with them to ensure
local charge neutrality), together with an extra variable because of
the new degree of freedom associated with the neutron superfluid, its
velocity ${\bf v}_n$, which we now define.

Pairing in a Fermi superfluid at point $\rr$ is described by the
amplitude $\langle {\hat \psi}_{ \uparrow}(\rr) {\hat \psi}_{ \downarrow}(\rr)
\rangle$, where ${\hat \psi}_{\sigma}(\rr)$ is the annihilation field
operator for a neutron with spin projection $\sigma$ at position $\rr$, and
$\langle\ldots\rangle$ denotes a quantum-mechanical or, at nonzero
temperature, a thermal average.  For a superfluid at rest, the
phase $2\varphi({\bf r})$ of this amplitude is independent of
position, while for a moving superfluid it depends on space. The
presence of the lattice of nuclei results in rapid spatial variations
on the scale of the nuclear spacing due to disturbances in the flow
produced by the nuclei, whereas in a translationally invariant system the
phase varies smoothly in space.  Just as for single-electron wave
functions in a periodic lattice, one may characterize states of the
neutron superfluid in the crust by a crystal momentum $\kk$ such that the
change of the phase $\varphi(\rr)$ when the spatial coordinate is
displaced by a lattice vector ${\bf R}$ is given by $\kk\cdot {\bf
  R}/\hbar$.  Alternatively, one may define a coarse-grained average phase,
$\phi(\rr)$, the average of $\varphi$ over a region in the vicinity of
$\rr$ with dimensions large compared with the lattice spacing but
small compared with other length scales in the problem, in which case
$\kk=\hbar \boldsymbol \nabla \phi$.  An analogous situation occurs for
atomic Bose--Einstein condensates in optical lattices, artificial
crystal lattices created by standing-wave laser beams
\cite{machholm}.

To derive the expression for the current density of neutrons, we
consider a system where there is no net current of the various
species.  Under a Galilean transformation to a reference frame moving
with velocity $-{\bf v}$ with respect to the original one, the wave
function of the nucleons is multiplied by a space-dependent factor
$\exp\left({\sum_j i m {\bf v}\cdot{\bf r}_j/\hbar}\right)$, where
the sum is over all nucleons.\footnote{For simplicity of exposition,
  we do not consider explicitly the electrons, which are
  relativistic.}  From the Galilean invariance of the system, the
neutron-number current density in the new frame is given by
\be
\jj_n=n_n \vv\,.
\label{neutroncurrent0}
\ee   
In this equation, both $\jj_n$ and $n_n$ refer to coarse-grained
average quantities, as described above for $\phi$.  On the other hand,
the current may be calculated directly in terms of the variation of
the phase of the pairing amplitude.  Under the Galilean
transformation, the coarse-grained phase $\phi$ of the pairing
amplitude, which is independent of space in the original frame,
acquires a space-dependent contribution $m \vv\cdot \rr/\hbar$, so one
can rewrite the expression (\ref{neutroncurrent0}) as
\be
\jj_n=n_n \vv_n\,,
\label{neutroncurrent1}
\ee
where 
\be
\vv_n=\frac{\hbar \boldsymbol \nabla \phi}{m}\,.
\label{vn}
\ee
As one can see from Eq.\ (\ref{neutroncurrent1}), the quantity $\vv_n$ is
the average neutron velocity in this situation and, quite
generally, it is the neutron superfluid velocity in a two-fluid model.

Now let us turn to the case where the nuclei are moving. In general, the velocity of the nuclei is given in terms of the displacement $\boldsymbol \xi$ of a nucleus by
\be
\vv_p=\frac{d {\boldsymbol \xi}}{dt}\,.
\ee 
Due to neutron--proton interactions, the motion of the protons induces a current of neutrons, an effect often referred to as ``entrainment''.  In the literature on quantum liquids it corresponds to ``backflow''.  We first consider the case where the phase $\phi$ is independent of space.  Phenomenologically one may write for the neutron current
\be
\jj_n= n_n^n (\vv_p)_0\,,
\ee
where the parameter $n_n^n$, which in general is a function of
$\vv_p$, corresponds to the density of neutrons that move with the
nuclei, and the subscript ``0'' on the velocity indicates that it is the velocity in the frame in which $\phi$ is independent of space.  In the context of the two-fluid model for superfluids, this
contribution corresponds to the {\it normal} component, which is why
we give it the superscript $n$.  
Because the solid is generally anisotropic due
to the presence of the crystal lattice, $n_n^n$ is a second-rank
tensor, but to simplify the discussion we shall treat it as a scalar.
For small values of $\vv_p$, we may take $n_n^n$ to be independent of
$\vv_p$.

Next we perform a Galilean transformation to the frame moving with velocity $-\vv$.  The neutron current in the new frame is given by
\be
\jj_n=n_n^n( \vv_p)_0+n_n\vv\,,
\label{neutroncurrent}
\ee 
which we now express in terms of velocities measured in the new frame.    The proton velocity  in the new frame is given by
\be
\vv_p=(\vv_p)_0+\vv
\ee
and the velocity $\vv_n$ by $\vv$ and therefore the neutron current density (\ref{neutroncurrent}) is given by
\be
\jj_n=   n_n \frac{\hbar { \boldsymbol \nabla} \phi}{m} +n_n^n(\vv_p-\vv_n)            =  n_n^n\vv_p+n_n^s\vv_n,
\label{neutroncurrent2}
\ee
where $n_n^s=n_n-n_n^n$ is referred to as the neutron superfluid
number density. The first equality in (\ref{neutroncurrent2}) is
analogous to the expression $\hat \pp -{e\bf A}$ for the current
operator for a charged particle in the presence of a vector potential,
where $\hat \pp$ is the momentum operator and $\bf A$ is the vector
  potential.  Thus we see that a difference between the velocities of
  the normal and superfluid components gives rise to a vector
  potential acting on the neutrons.

Equation (\ref{neutroncurrent2}) provides the foundation for a
two-fluid description of motion of matter in the inner crust.  We
shall assume that frequencies of interest are much less than the
electron plasma frequency, in which case the electron density is equal
to the proton density.  The long-wavelength dynamics of the system may
thus be described in terms of the coarse-grained averages of the
densities of neutrons and protons, $n_n$ and $n_p$, and the velocities
$\vv_n$ and $\vv_p$.  On time scales long compared with the time for
weak interaction processes, the numbers of neutrons and protons are
separately conserved.  The equation for neutron number conservation is
\be
\frac{\partial n_n}{\partial t} +{ \boldsymbol \nabla}\cdot \jj_n=\frac{\partial n_n}{\partial t} +{ \boldsymbol \nabla}\cdot (n_n^n\vv_p+n_n^s\vv_n)  =0
\label{neutronnumber}
\ee
and that for proton number conservation is
\be
\frac{\partial n_p}{\partial t} +{ \boldsymbol \nabla}\cdot \jj_p=0,
\label{protonnumber}
\ee
where the proton current density is given by
\be
\jj_p=n_p\vv_p.
\ee
These must be supplemented by equations for the time derivatives of $\vv_n$ and $\vv_p$.  The rate of change of the phase $\phi$ is given by the Josephson relation 
\be
\frac{\partial \phi}{\partial t}=- \frac{\mu_n}{\hbar},
\label{josephson}
\ee
where $\mu_n$ is the neutron chemical potential, which is related to the energy per unit volume $E$ by the equation $\mu_n=\partial E/\partial n_n$. \footnote{In Secs.\ \ref{sec:hydrodynamics} and \ref{sec:collective modes}, $E$ denotes the energy per unit volume, not the energy.}
Thus, on taking the gradient of Eq.\ (\ref{josephson}) one finds
\be
m\frac{\partial \vv_n}{\partial t}+ { \boldsymbol \nabla}\mu_n=0,
\label{dvndt}
\ee

The final equation is that for momentum conservation.  The total momentum density $\bf g$ is the sum of the neutron and proton momentum densities,
\be
{\bf g}=m\jj_n+m\jj_p,
\ee
where we have neglected the contribution of the electrons.  The latter is of order $\mu_e/mc^2$ times the proton contribution  and the electron chemical potential, $\mu_e$, is less than 10\% of the proton rest mass. Conservation of momentum is expressed by the relation 
\be
\frac{\partial {\bf g}}{\partial t} +{ \boldsymbol \nabla}\cdot {\textbf{\textsf P}}=0,
\label{momcons}
\ee
where $ \textbf{\textsf P}$  is the momentum flux (stress) tensor.

\subsection{Long-wavelength collective modes}
\label{sec:collective modes}
As an application of the above equations, we consider long-wavelength, small-amplitude  oscillations of the medium about a state in which the velocities are zero.  For simplicity we consider a uniform medium that is homogeneous initially.  We  linearize Eqs.\ (\ref{neutronnumber}), (\ref{protonnumber}), (\ref{dvndt}), and (\ref{momcons}).  To lowest order in the velocities the first two of these equations become
\be
\frac{\partial n_n}{\partial t}  +n_n^n{ \boldsymbol \nabla}\cdot \vv_p+n_n^s{ \boldsymbol \nabla}\cdot \vv_n  =0\,,
\label{neutronnumberlin}
\ee
and
\be
\frac{\partial n_p}{\partial t}  +n_p{ \boldsymbol \nabla}\cdot \vv_p =0\,.
\label{protonnumberlin}
\ee
The neutron chemical potential is a function of the neutron and proton densities, and therefore changes of it are given by
\be
\delta \mu_n=E_{nn}\delta n_n +E_{np}\delta n_p\,,
\ee
 where $E_{ij}=\partial \mu_i/\partial n_j=\partial^2E/\partial n_i\partial n_j$, and Eq.\ (\ref{dvndt}) becomes
\be
m\frac{\partial \vv_n}{\partial t}+ E_{nn}{ \boldsymbol \nabla}n_n + E_{np}{ \boldsymbol \nabla}n_p=0\,.
\label{dvndt2}
\ee

In the case of a solid, the momentum flux tensor contains contributions from elastic forces and to describe these it is convenient to use the nuclear displacement, $\boldsymbol \xi$, rather than the velocity $\vv_p$.  The change in the proton density is given by
\be
\delta n_p=-n_p {\boldsymbol \nabla}\cdot {\boldsymbol \xi}\,.
\ee   

The momentum flux tensor has contributions from elastic distortions of the lattice as well as from changes in the neutron density.  In neutron star crusts, matter is expected to be microcrystalline, and at wavelengths longer than the typical size of the single-crystal domain, the medium behaves elastically as a uniform medium and the changes in the momentum flux tensor  are given by
\be
\delta {{\textsf P}}_{ij}=(n_n\delta \mu_n +n_p \delta \mu_p)\delta_{ij} - S \left(\frac{\partial  \xi_i}{\partial x_j}+\frac{\partial  \xi_j}{\partial x_i}-\frac{2\delta_{ij}}{3}\sum_{k=1}^3 \frac{\partial  \xi_k}{\partial x_k}\right)\,, 
\ee
where $S$ is the effective shear elastic constant.  

Detailed calculations of collective modes have been carried out in a number of works \cite{reddymodes, chamelmodes, kobyakov1} and here we shall summarize the results.  For an isotropic medium, there are two transverse modes, corresponding to lattice phonons with two orthogonal polarizations, and two longitudinal modes.
In the absence of neutron-proton interactions, there would be no entrainment, and the transverse phonons would have a velocity \footnote{To realize the situation envisaged here, one must imagine that when the neutron--proton interaction is absent,   the proton--proton interaction is increased so that pure proton nuclei are bound.}
\be
v_t=\sqrt{\frac{S}{mn_p}}\,,\,\,\, \text{ (No entrainment)}
\ee   
while with neutron--proton interactions included the result is 
\be
v_t=\sqrt{\frac{S}{m(n_p+n_n^n)}}\,.  \,\,\, \text{ (Superfluid neutrons)}\
\ee  
This result reflects the fact that the superfluid flow must be irrotational, ${ \boldsymbol \nabla}\times \vv_n=0$, and consequently only the normal density of the neutrons enters.  By contrast,
when the neutrons are in the normal state, all of them contribute to the relevant mass density and
 \be
v_t=\sqrt{\frac{S}{\rho}}\,,  \,\,\,  \text{ (Normal neutrons)}\
\ee  
where $\rho$ is the total mass density, which is approximately 
$m(n_p+n_n)$ in the nonrelativistic limit.
In the absence of neutron--proton interactions, the two longitudinal modes correspond to phonons in the lattice and phonons in the neutron superfluid (the Bogoliubov--Anderson mode) and they have velocities
\be
v_{p}=\sqrt{\frac{n_p^2E_{pp}+\frac{4}{3} S}{mn_p}}\,,\,\,\, \text{ (No neutron-proton interaction)}\
\ee  
 and 
 \be
v_{n}=\sqrt{\frac{n_n E_{nn}}{m}}\,,\,\,\, \text{ (No neutron-proton interaction)}.
\ee  
 
When entrainment is included but coupling between the nucleon densities in the two modes is neglected, the mass density associated with the lattice phonons increases and that associated with the phonons in the neutron superfluid decreases, thereby decreasing the velocities of both modes:
\be
v_{p}=\sqrt{\frac{{\cal E}^{nn}+\frac43 S}{m(n_p+n_n^n)}}\,,\,\,\, \text{ (With entrainment but no hybridization)}
\ee   
and 
\be
v_{n}=\sqrt{\frac{n_n^s E_{nn}}{m}}\,,\,\,\, \text{ (With entrainment but no hybridization)}
\ee 
where
\be
{\cal E}^{nn}=n_p^2E_{pp} +2 n_p n_n^nE_{np}+(n_n^n)^2 E_{nn}\,.
\ee

Finally, when interactions between the densities of neutrons and protons in the two modes are included, the modes are hybridized and their velocities are given by
\be
v_{\pm}^2=\frac{v_n^2+v_p^2}2 \pm\sqrt{\left(\frac{v_n^2-v_p^2}{2}\right)^2+  v_{np}^4}\,\,\,,
\label{vpm}
\ee
where
\be
 v_{np}^2=\left(\frac{n_n^s}{n_n^n+n_p}\right)^{1/2}  \frac{  E_{nn}n_n^n+E_{np}n_p}{m}\,.
 \label{vnp}
\ee

\begin{figure}
\begin{center}
\includegraphics*[width=1.0\columnwidth]{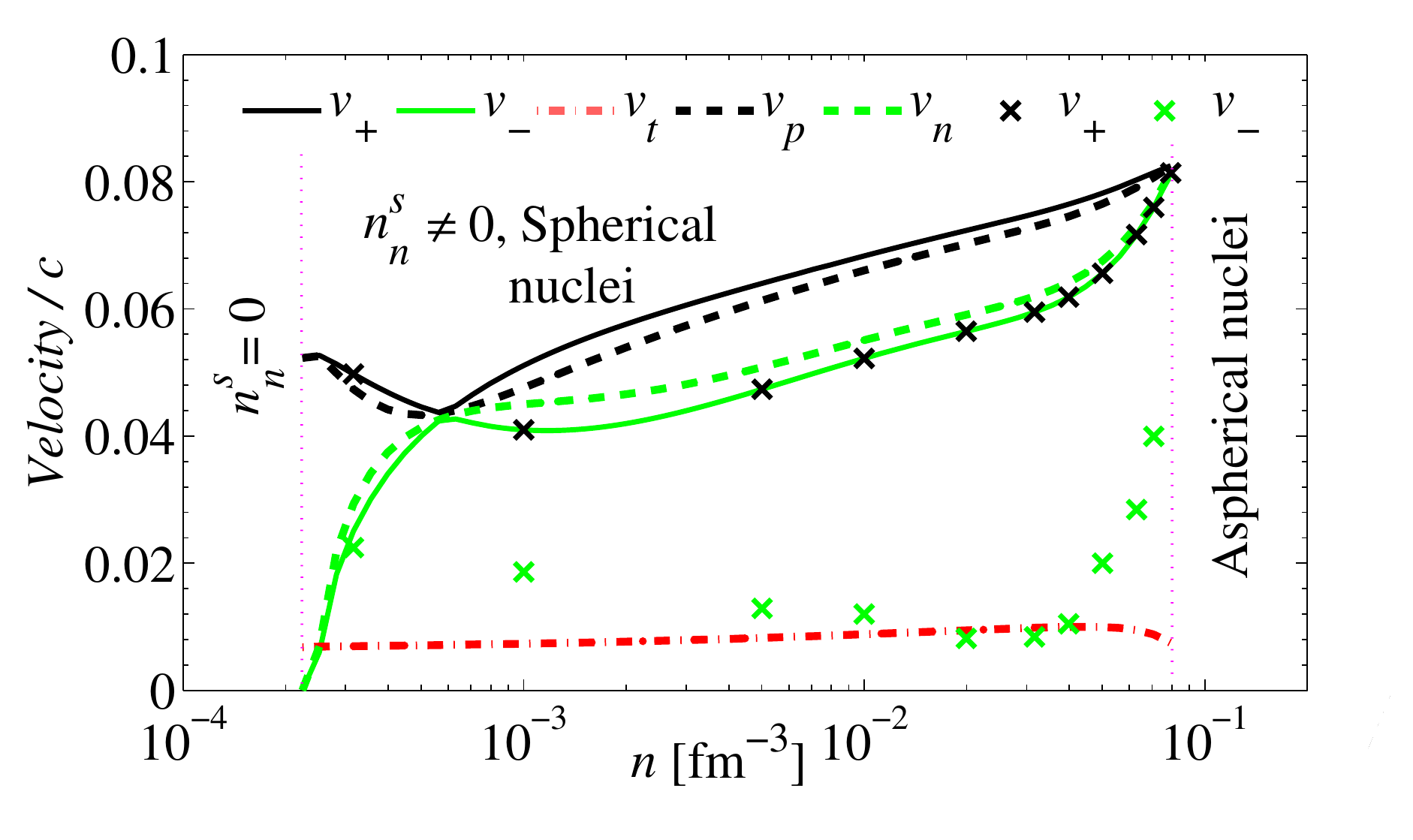}
\caption{Velocities of long-wavelength modes as a
function of nucleon density \cite{kobyakov1}. Velocities of modes when effects of entrainment are included are given
by $v_n$ and $v_p$ when $v_{np}$ is neglected, and  by
$v_\pm$  when $v_{np}$ is included. The velocity of the transverse modes is denoted by  $v_t$. Curves are for
$n_n^s
= n^{\rm out}$, the local density of neutrons outside nuclei, while the crosses show results for $n_n^s$
taken from Chamel's
calculation \cite{chamel2012}.  The derivatives $E_{ij}$ were extracted from Lattimer and Swesty's equation of state calculations \cite{lattimer}. } \label{fig:modefreqs}
\end{center}
\end{figure}

When the neutrons are normal, there is a single longitudinal mode, a sound wave, with velocity
\be
v_{\rm sound}= \sqrt{\frac{B+\frac43S}{\rho}}\,,
\ee
where $B$ is the bulk modulus,
\be
B=n_n^2E_{nn}+2 n_n n_p E_{np}+n_p^2E_{pp}\,.
\ee

In Fig.~\ref{fig:modefreqs} we give an example of a calculation of
mode frequencies.  As one sees from the figure, these depend rather
strongly on the value of the neutron superfluid density, a topic which
we take up in the following subsection.

Recently, the calculations of mode frequencies have been extended to
shorter wavelengths~\cite{kobyakov2}.  An interesting conclusion of
that work is that the body-centered-cubic lattice of ions, which has
generally been assumed to be the ground state because it has the
lowest Coulomb lattice energy, may be destabilized by the presence of
neutrons between nuclei.  The physical origin of the effect is that
the neutrons between nuclei give rise to an attractive induced
interaction between nuclei which tends to favor inhomogeneous states
in which regions of a high density of nuclei also have a high density
of neutrons.  This attraction is similar to the phonon-induced
attraction between electrons that is responsible for superconductivity
in metals and to the induced interactions between neutrons discussed in
Sec.\ \ref{sec:inducedinteractions}.  The attraction is insufficient
to cause instability at long wavelengths, but could be large enough to
do so at shorter wavelengths where the other part of the
nucleus--nucleus interaction, a Coulomb interaction screened by
electrons, is reduced.  We note that, while in the calculations of
mode frequencies we have assumed that the neutrons are superfluid, 
the induced interaction would also be present if the neutrons are
normal.

\subsection{Band structure and the neutron superfluid density}

Since the neutrons are subject to the periodic potential produced by
the nuclei, the neutron spectra display a band structure.  Even in the
normal state, calculating the band structure of the system is very
difficult, because the number of bands that are significantly occupied
is of the order of half the number of neutrons per unit cell, the half
being due to the spin degeneracy of the neutrons.  This can be of
order $10^3$, whereas in terrestrial condensed matter the number of
bands considered for simple crystal structures is typically two orders of magnitude less.  
The large number of bands places extreme demands on
the numerical accuracy required.  The difficulty of performing
band-structure calculations motivated the use of the Wigner-Seitz
approximation described in Sec.\ \ref{sec:staticproperties}.  However,
this approximation is inadequate to describe the effects of band
structure on flow properties such as the neutron superfluid density,
which is an important quantity for mode frequencies and for models of
glitch phenomena based on a neutron superfluid moving relative to the
lattice of nuclei in the crust: in both these cases the total moment
of inertia of the superfluid neutrons is a crucial parameter.  The
difficulty in calculating the neutron superfluid density stems from the fact
that the neutrons move in the periodic lattice of nuclei.  Were it not
for the lattice, the neutron superfluid density would be equal to the
total density of neutrons.  However, the medium is highly
inhomogeneous on a microscopic scale, with nuclei containing nuclear
matter with both protons and neutrons, and a neutron fluid between
nuclei.  Physically, one might expect that a good estimate of the
neutron superfluid density would be the density of neutrons in the
regions outside nuclei, $n^{\rm out}$.  However, as will be explained
below, Chamel has performed band-structure calculations for neutrons
in a lattice of nuclei, and finds the superfluid density to be an
order of magnitude smaller \cite{chamel2012}.  We now describe some of
the issues relevant to calculating the superfluid density.

The quantity $n_n^s$ represents the linear response of the current to a change in the gradient of the phase of the pairing amplitude.  Let us begin by considering 
noninteracting particles moving in a periodic potential.    The single-particle states are specified by a crystal momentum $\kk$ and a band index $b$, and the current density carried by the particles is\footnote{Note that, in this subsection, $\kk$ is a crystal momentum, not a wave vector.}
\be
\jj_n=\frac1\Omega\sum_{\kk, b}{f_{\kk,b}}\vv_{\kk,b}\,,
\ee
where $f_{\kk,b}$ is the particle distribution function, $\Omega$ is the volume of the system, and the current carried by a particle is its velocity, 
\be
\vv_{\kk,b}={\boldsymbol \nabla}_{\kk} \epsilon_{\kk,b}\,.
\ee
 Here  $ \epsilon_{\kk,b}$ is the energy of the state with crystal momentum $\kk$ and band index  $b$, and the sum is over all states within the first Brillouin zone.\footnote{For simplicity, we do not indicate explicitly spin indices, which must also be summed over.}    When an extra phase $\chi$ is applied to the single-particle states, the wave vector of the state is changed by an amount ${\boldsymbol \nabla}\chi$.  On the assumption that the change of the phase can be made adiabatically, the distribution function for the state $\kk$ after addition of the phase change is equal to that for the state $\kk-\qq$ initially, where $\qq=\hbar {\boldsymbol \nabla}\chi$ and the current density is given by
 \be
 \jj_n=\frac1\Omega\sum_{\kk, b}{f_{\kk-\qq,b}}\vv_{\kk,b}\simeq -\frac1\Omega\sum_{\kk, b}\vv_{\kk,b}\,\,\qq\cdot {\boldsymbol \nabla}_{\kk} f_{\kk,b}\,,
 \ee
where we have assumed that the current vanishes in the initial state and the second expression holds for small $\qq$.  On writing $ {\boldsymbol \nabla}_{\kk} f_{\kk,b}=  (\partial f_{\kk,b}/\partial \epsilon_{\kk,b}){\boldsymbol \nabla}_{\kk}  \epsilon_{\kk,b}=(\partial f_{\kk,b}/\partial \epsilon_{\kk,b})\vv_{\kk,b}$ one finds
\be
 \jj_n\simeq -\frac1\Omega\sum_{\kk, b}\vv_{\kk,b}\,(\qq\cdot \vv_{\kk,b}) \frac{\partial f_{\kk,b}}{\partial \epsilon_{\kk,b}}=  \frac{\qq}\Omega\sum_{\kk, b}\frac{v_{\kk,b}^2}3\,\left(-\frac{\partial f_{\kk,b}}{\partial \epsilon_{\kk,b}}\right) \, ,
 \label{current3}
 \ee
where the latter expression holds for isotropic matter and for crystals with cubic symmetry.\footnote{For more complicated crystal structures the current is not generally in the direction of $\qq$ and the response is described by a tensor.} The quantity  
\be
\frac{1}\Omega\sum_{\kk, b}\frac{v_{\kk,b}^2}3\,\left(-\frac{\partial f_{\kk,b}}{\partial \epsilon_{\kk,b}}\right) =\frac{1}{\Omega}\sum_{\kk, b}\frac{f_{\kk,b}}{3}\nabla_\kk^2\epsilon_{\kk,b} 
\ee
has the dimensions of a number density divided by a mass.  If one defines an effective mass by the relationship $1/m^{\rm eff}_{\kk,b}=1/3 \sum_i \partial^2\epsilon_{\kk,b}/\partial k_i^2 $, one sees that the quantity is the neutron density divided by the harmonic mean of the effective mass for occupied states.   Fully occupied bands do not contribute to this quantity because of the periodicity of $\epsilon_{\kk,b}$ in $\kk$ space. 

If neutrons are not superfluid, impurities in the lattice will transfer momentum to the lattice  and the assumption that the neutrons adjust adiabatically to an imposed phase change will not hold.  However, for a superfluid the assumption will be good, since pairing correlations maintain the coherent motion of the neutrons.  To the extent that energies associated with pairing are small compared with other energies in the problem, such as the Fermi energy and band gaps, the response of the neutrons should be given approximately by the result (\ref{current3}).  In that case, the neutron superfluid velocity is given by
\be
\vv_n=\frac{\qq}{m}\,,
\ee 
and therefore one sees that
\be
n_n^s=\frac{m}\Omega\sum_{\kk, b}\frac{v_{\kk,b}^2}3\,\left(-\frac{\partial f_{\kk,b}}{\partial \epsilon_{\kk,b}}\right)\,. 
\ee
On replacing the distribution function by the zero temperature form $\Theta(\mu_n-\epsilon_\kk)$, where $\Theta$ is the Heaviside step function, one finds
\be
n_n^s =\frac13\int_{FS} \frac{   mv_{\kk,b}   \cdot d{\bf S}_\kk}{(2\pi\hbar)^3}  \,,
\label{nnsBand}
\ee
where the integral is to be carried out over the Fermi surface.  The quantity $d{\bf S}_\kk$ is an element of the area of the Fermi surface. For a spectrum $\epsilon_\kk=k^2/2m_B$, where $m_B$ is a parameter, Eq.\ (\ref{nnsBand}) gives $n_n^s=(m/m_B)n_n$. 
If one writes the expression for the neutron current in the form $\jj_n= \Xi \,\qq$, the response function $\Xi$ has the dimensions of a density divided by a mass.  If one chooses the mass to be the bare neutron mass, one may regard a reduction of $\Xi$ due to the periodic lattice as a reduction in the effective number of free neutrons.  Alternatively, if one chooses some value for the density, a reduction of $\Xi$ can be attributed to an increase in the effective mass of neutrons.  To obtain a treatment that is close to the conventional one for the two-fluid model \cite{LLHydro}, it is convenient to choose as the mass the bare neutron mass and regard any reduction in $\Xi$ as being a reduction of the number of superfluid neutrons compared with the total number.  This choice of mass has the advantage that the expression for the current, Eq.\ ({\ref{neutroncurrent2}}), naturally involves the difference $\vv_p-\vv_n$: had we chosen a different mass $\tilde m$ in the definition of the neutron velocity, Eq. (\ref{vn}), the uglier quantity $({\tilde m}/m)\,\vv_p-\vv_n$  would appear instead.  Of course, no physical results depend on the choice of the mass in the definition of the superfluid velocity, since the basic physical quantity of interest is the superfluid momentum per particle, $\hbar {\boldsymbol\nabla}\phi$, as has been stressed in Ref.\ \cite{chamel_haensel}.  The traditional treatment brings out directly the fact that there is a single parameter, $n_n^s$, that describes the flow of the superfluid, whereas if one works in terms of momenta per particle,  the currents are related to the momenta per particle by a tensor.

\begin{figure}
\begin{center}
\includegraphics*[width=0.7\columnwidth]{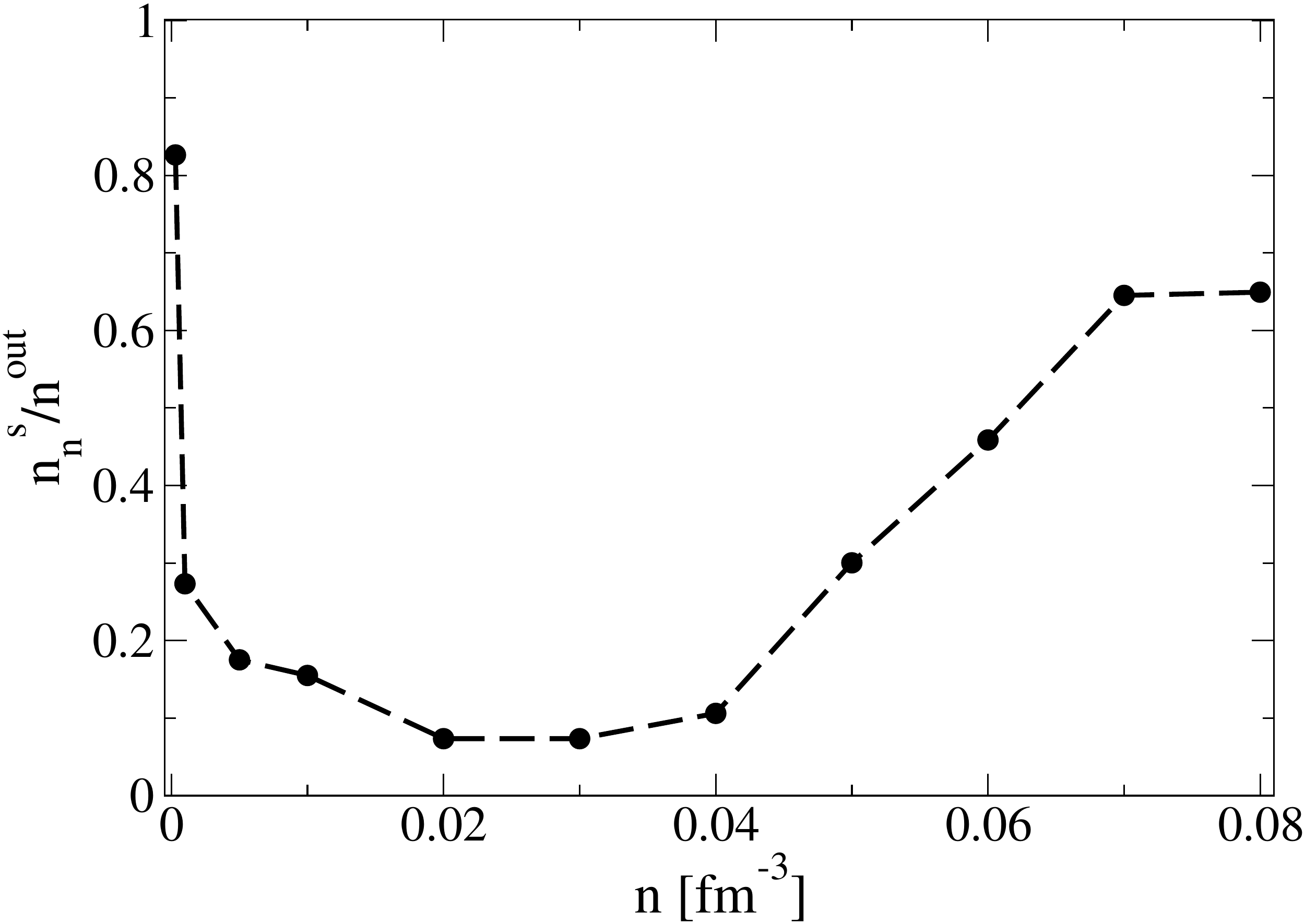}
\caption{Neutron superfluid density $n_n^s$ in terms of the density of neutrons outside nuclei, $n^{\rm out}$, as a function of the total nucleon \mbox{density $n$ \cite{chamel2012}.}} \label{fig:chamel_nns}
\end{center}
\end{figure}

We now turn to Chamel's calculations \cite{chamel2012}.  He calculated in the Hartree--Fock approximation the band structure of neutrons moving in 
a body-centered-cubic lattice of nuclei.  In Fig.\ \ref{fig:chamel_nns} we show Chamel's results for  $n_n^s/n^{\rm out}$ as a function of the total density of matter.
Are there other physical effects which could change this result significantly?  A basic assumption in using the result (\ref{nnsBand})
is that pairing correlations have only a small effect on the neutron excitations.  This is presumably a reasonable assumption provided the pairing gap $\Delta $ is small compared with splittings between  single-neutron states produced by the periodic lattice.
We know of no published numerical values of these splittings, but plots of neutron spectra are shown in Figs.\ 2-4 of Ref.~\cite{chamel2012}. At a density of 0.0003 fm$^{-3}$, which corresponds to a mass density $\sim$ 5$\times 10^{11}$\,g cm$^{-3}$, slightly larger than the neutron drip density, the bands are very similar to those of free neutrons and splittings are of order 0.1 MeV.    At a density 0.03 fm$^{-3}$ (Fig.\ 3 of Ref.~\cite{chamel2012}), the band structure is complicated and it is difficult to resolve the splittings, while at a density of  0.08 fm$^{-3}$, close to the inner edge of the crust, the bands are again free-neutron like and splittings are of order 1 MeV.    Pairing gaps in neutron matter at densities of order $n_s/10$ are typically of order 1 MeV which is similar in magnitude to splittings due to band structure: this indicates that pairing could have a large effect on the band structure.  Pairing reduces the effects of band structure because the pairing gap increases the size of energy denominators for processes in which neutron quasiparticles are scattered by the periodic lattice.  

\begin{figure}
\begin{center}
\includegraphics*[width=0.5\columnwidth]{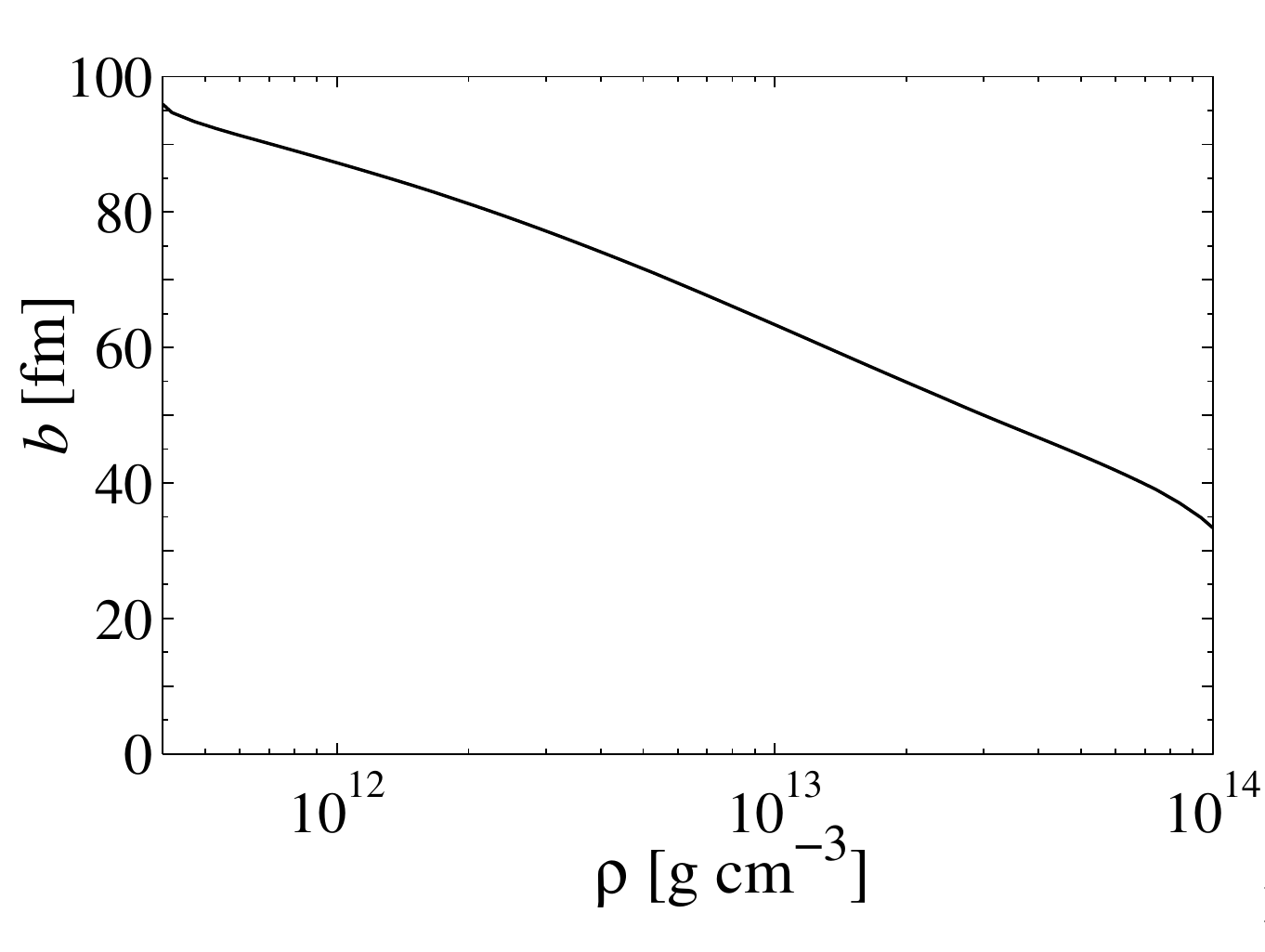}
\caption{Lattice spacing $b$ of the body-centered-cubic crystal in the inner crust as a function of the total matter density (from Ref.\ \cite{kobyakov2}, which was based on the calculations of Ref.\ \cite{lattimer}).}
\label{fig:latticespacing}
\end{center}
\end{figure}

Another useful way to assess the importance of various effects is to estimate basic length scales.  One important length is the separation between nuclei, and we show in Fig.\ \ref{fig:latticespacing} a plot of the lattice spacing of the body-centered-cubic cell in the neutron star crust as a function of the total density of nucleons.   Another relevant length is the coherence length for the pairing correlations, $\xi_{\rm coh}$, which is a measure of the size of a Cooper pair.  In BCS theory this is given for uniform matter by 
\be
\xi_{\rm coh}=\frac{\hbar v_F}{\pi \Delta} \approx \frac{63\,\, {\rm MeV}}{\Delta}\frac{v_F}{c}\,\,\,{\rm fm}  \, ,
\ee  
where $v_F$ is the Fermi velocity.  Thus, in the density range where the pairing gap for neutrons reaches its maximum value ($\sim 1$ MeV), the coherence  length is of order 10 fm, which is considerably less than the lattice spacing, while for neutron densities for which the gap is of order 0.1 MeV or less, it becomes larger than the lattice spacing.  These estimates suggest that in the density range where the calculations of $n_n^s$ based on the normal state band structure predict a strong reduction, correlated pairs of neutrons have a size much less than the lattice spacing and therefore neutron excitations propagate between scatterings from nuclei in the lattice not as free particles, but as the elementary excitations of the neutron superfluid. 

Another effect that could play a role is the zero-point and thermal motion of the lattice, which will smear out in space the potential due to nuclei and thereby reduce scattering from the lattice.  The smearing is described by the Debye--Waller factor and preliminary estimates indicate that the effect could be important \cite{kobyakov1}.

\section{Conclusion} 

During the past two decades, considerable progress has been made in
understanding $^1$S$_0$ pairing in neutron matter.  The impetus for
these advances has come from a number of sources.  One is that the
experimental realization of degenerate gases of fermionic atoms led to
the rediscovery of the work of Gor'kov and Melik-Barkhudarov
\cite{Gor'kov}, which showed that induced interactions play an
important role even in the limit of low density.  Their analytical
results, together with the development of quantum Monte Carlo methods
for solving the many-body problem have resulted in a reliable,
quantitative understanding of gaps in neutron matter at low densities.

At higher densities there are significant uncertainties in the gap
because, with increasing density, non-central components of 
nucleon--nucleon interactions become more important, thereby making the
many-body problem more difficult to solve.  In addition, three-and
higher-body forces come into play.  It is, for example, difficult to
predict the density at which the neutron $^1$S$_0$ gap vanishes.  In
the future, one may anticipate that the development of quantum Monte Carlo
methods, as well as of improved effective interactions, will make for
more precise estimates.  Even more uncertain than the $^1$S$_0$ gap
are estimates for the $^3$P$_2$ gap (if any) because this, unlike the
$^1$S$_0$ gap, which depends on the average of the pairing interaction
over the Fermi surface, depends on ${\it departures}$ of the pairing
interaction from its average value.  In addition, $^3$P$_2$ is a
viable pairing channel only at densities around nuclear
saturation density or above, where the many-body problem is difficult to solve and nuclear interactions are less constrained.
Predicting pairing gaps for protons is bedevilled by the fact that the
protons are a minority component in neutron star matter, and
consequently induced interactions due to the surrounding neutrons play
a large, but poorly understood, role.

Long-wavelength phenomena may be described in terms of a two-fluid
model similar to the standard one for the helium liquids, but
generalized to include the elastic effects of the crystal lattice.
With improved effective interactions to describe nuclear properties,
the road is now open to better determinations of the parameters
entering the two-fluid model.  Of particular importance for improved
calculations of mode frequencies is a better understanding of the
neutron superfluid density, and here a deeper study of the combined
effects of pairing and band structure is called for.  Properties of
collective modes at shorter wavelengths are of interest for thermal
and transport properties, but to date little work has been done on
this problem.

\section*{Acknowledgments}

We thank Joe Carlson, Nicolas Chamel, Kai Hebeler, Dmitry Kobyakov, and Alessandro
Pastore for helpful input and discussions. This work was supported in
part by the Natural Sciences and Engineering Research Council of
Canada, the ERC Grant No.~307986 STRONGINT, the Helmholtz
Association through the Helmholtz Alliance Program, contract
HA216/EMMI ``Extremes of Density and Temperature: Cosmic Matter in the
Laboratory'', and the NewCompStar network, COST Action MP1304.


\addcontentsline{toc}{section}{References}
\begin{thebibliography}{99}
\bibitem{page_this volume} Page, D., Lattimer, J.\ M., Prakash, M., 
and Steiner, A.\ W.\ in
{\it Novel Superfluids: Volume 2}. Eds. Bennemann, K.\ H.\ and
Ketterson, J.\ B., Oxford University Press, Oxford (2014), p.~505 [arXiv:1302.6626].

\bibitem{LombardoSchulze} Lombardo, U.\ and Schulze, H.-J.\
Superfluidity in neutron star matter. In {\it Physics of Neutron Star
Interiors}, eds.\ Blaschke, D., Glendenning, N.\ K., and Sedrakian, A.\ 
Lecture Notes in Physics {\bf 578}, 30 (2001).

\bibitem{Dean:2002zx} Dean, D.\ J.\ and Hjorth-Jensen, M.\
Pairing in nuclear systems: From neutron stars to finite nuclei,
{\it Rev.\ Mod.\ Phys.} {\bf 75}, 607 (2003).


\bibitem{Bohr:1958} Bohr, A., Mottelson, B.\ R., and Pines, D.\ 
Possible analogy between the excitation spectra of nuclei and those
of the superconducting metallic state. {\it Phys.\ Rev.} {\bf 110}, 
936 (1958).

\bibitem{BrinkBroglia} Brink, D.\ M.\ and Broglia, R.\ A.\ 
{\it Nuclear Superfluidity: Pairing in Finite Systems}.
Oxford University Press, Oxford (2005).

\bibitem{migdal} Migdal, A.\ B.\ Superfluidity and the moments of 
inertia of nuclei. \textit{Zh.\ Eksp.\ Teor.\ Fiz.} {\bf 37} 249 (1959)
[\textit{Sov.\ Phys.\ JETP} {\bf 10}, 176 (1960)]; \textit{Nucl.\ Phys.}
{\bf 13}, 655 (1959).

\bibitem{ginzburgkirzhnits} Ginzburg, V.\ L.\ and Kirzhnits, D.\ A.\
On the superfluidity of neutron stars. \textit{Zh.\ Eksp.\ Teor.\ 
Fiz.} {\bf 47}, 2006 (1964) [\textit{Sov.\ Phys.\ JETP} {\bf 20}, 
1346 (1965)].

\bibitem{ginzburg} Ginzburg, V.\ L.\ Superfluidity and superconductivity
in the Universe. \textit{Usp.\ Fiz.\ Nauk} {\bf 97}, 601 (1969) 
[\textit{Sov.\ Phys.-Uspekhi} {\bf 12}, 241 (1969)]; 
\textit{J.\ Stat.\ Phys.} {\bf 1}, 3 (1969).

\bibitem{bpp} Baym, G., Pethick, C.\ J., and Pines, D.\ Superfluidity
in neutron stars. \textit{Nature} {\bf 224}, 673 (1969).

\bibitem{hoffberg} Hoffberg, M., Glassgold, A.\ E., Richardson, R.\ W.,
and Ruderman, M.\ Anisotropic superfluidity in neutron star matter. 
\textit{Phys.\ Rev.\ Lett.} {\bf 24}, 775 (1970).

\bibitem{Stoks:1993} Stoks, V.\ G.\ J., Klomp, R.\ A.\ M., Rentmeester, 
M.\ C.\ M., and de Swart, J.\ J.\ Partial-wave analysis of all 
nucleon-nucleon scattering data below 350~MeV. {\it Phys.\ Rev.\ C}, 
{\bf 48}, 792 (1993).


\bibitem{clark} Clark, J.\ W., Dav{\'e}, R.\ D., and Chen, J.\ M.\ C.\ 
Nucleonic superfluids. In {\it Condensed Matter Theories, Vol.\ 8}, 
Plenum, New York (1993); see also Chen, J.\ M.\ C., Clark, J.\ W., 
Dav{}\'e, R.\ D., and Khodel, V.\ V.\ Pairing gaps in nucleonic 
superfluids. \textit{Nucl.\ Phys.\ A} {\bf 555}, 59 (1993).

\bibitem{Wambach:1993} Wambach, J., Ainsworth, T.\ L., and Pines, D.\
Quasiparticle interactions in neutron matter for applications in neutron
stars. \textit{Nucl.\ Phys.\ A} {\bf 555}, 128 (1993).

\bibitem{Degennes:1966} de Gennes, P.\ G.\ \textit{Superconductivity
of Metals and Alloys}. Westview Press, Boulder (1966).

\bibitem{Tinkham:1996} Tinkham, M.\ {\it Introduction to Superconductivity}.
McGraw-Hill, New York (1996).

\bibitem{Wiringa:1995} Wiringa, R.\ B., Stoks, V.\ G.\ J., and Schiavilla,
R.\ Accurate nucleon-nucleon potential with charge-independence breaking.
{\it Phys.\ Rev.\ C} {\bf 51}, 38 (1995).

\bibitem{Khodel:1996} Khodel, V.\ A., Khodel, V.\ V., and Clark, J.\ W.
Solution of the gap equation in neutron matter. {\it Nucl.\ Phys.\ A}
{\bf 598}, 390 (1996).

\bibitem{Gezerlis2008} Gezerlis, A. and Carlson, J. 
Strongly paired fermions: Cold atoms and neutron matter.
\textit{Phys. Rev. C} {\bf 77}, 032801(R) (2008); 
Gezerlis, A. and Carlson, J.
Low-density neutron matter.
\textit{Phys. Rev. C} {\bf 81}, 025803 (2010). 

\bibitem{ColdFermions}  For reviews, see Bloch, I., Dalibard, J., and Zwerger, W.
Many-body physics with ultracold gases.
\textit{Rev. Mod. Phys.} {\bf 80}, 885 (2008); 
Giorgini, S., Pitaevskii, L. P., and Stringari, S. 
Theory of ultracold atomic Fermi gases. 
\textit{Rev. Mod. Phys.} {\bf 80}, 1215 (2008).

\bibitem{Gor'kov} Gor'kov, L.\ P.\ and Melik-Barkhudarov, T.\ K.\ 
Contribution to the theory of superfluidity in an imperfect Fermi 
gas. {\it Zh.\ Eksp.\ Teor.\ Fiz.\/} {\bf 40}, 1452 (1961) 
[{\it Sov.\ Phys.\ JETP} {\bf 13,} 1018 (1961)].

\bibitem{HeiselbergPSV} Heiselberg, H., Pethick, C.\ J., Smith, H.,
and Viverit, L.\ Influence of induced interactions on the superfluid
transition in dilute Fermi gases. {\it Phys.\ Rev.\ Lett.} {\bf 85},
2418 (2000).

\bibitem{berkschrieffer} Berk, N.\ F.\ and Schrieffer, J.\ R.\ 
Effect of ferromagnetic spin correlations on superconductivity.
{\it Phys.\ Rev.\ Lett.} {\bf 17}, 433 (1966).

\bibitem{andersonbrinkman} Anderson, P.\ W.\ and Brinkman, W.\ F.\ 
Anisotropic superfluidity in $^3$He: A possible interpretation of 
its stability as a spin-fluctuation effect. {\it Phys.\ Rev.\ Lett.}
{\bf 30}, 1108 (1973).

\bibitem{martikainen} Martikainen, J.-P., Kinnunen, J.\ J., 
T{\"o}rm{\"a}, P., and Pethick, C.~J.\ Induced interactions and 
the superfluid transition temperature in a three-component Fermi gas.
{\it Phys.\ Rev.\ Lett.} {\bf 103}, 260403 (2009).

\bibitem{bertsch} Bertsch, G.\ F.\ \textit{The Many-Body Challenge 
Problem (MBX)}. See Bishop, R.\ F.\ \textit{Int.\ J.\ Mod.\ Phys.\ B}
{\bf 15}, 10, iii (2001).

\bibitem{unitarygap} Carlson, J., Gandolfi, S., and Gezerlis, A.\
Quantum Monte Carlo approaches to nuclear and atomic physics.
\textit{Prog.\ Theor.\ Exp.\ Phys.} 01A209 (2012).

\bibitem{PethickSmith} See, e.g., Pethick, C.\ J.\ and Smith, H.\ 
\textit{Bose--Einstein Condensation in Dilute Gases}, 2nd ed., 
Cambridge University Press, Cambridge (2008), Sec.\ 17.3.1.

\bibitem{SchwenkPethick}  Schwenk, A. and Pethick, C. J. Resonant Fermi gases with a large effective range
{\it Phys. Rev. Lett.} {\bf 95}, 160401 (2005). 

\bibitem{Bogner:2003wn} Bogner, S.\ K., Kuo, T.\ T.\ S., and Schwenk,
A.\ Model independent low momentum nucleon interaction from phase shift
equivalence. {\it Phys.\ Rept.}  {\bf 386}, 1 (2003).

\bibitem{Hebeler:2006kz} Hebeler, K., Schwenk, A., and Friman, B.\
Dependence of the $^1$S$_0$ superfluid pairing gap on nuclear 
interactions. {\it Phys.\ Lett.\ B} {\bf 648}, 176 (2007).

\bibitem{Hebeler:2009iv} Hebeler, K., and Schwenk, A.\ Chiral 
three-nucleon forces and neutron matter. {\it Phys.\ Rev.\ C}
{\bf 82}, 014314 (2010).

\bibitem{Zuo:2004mc} Zuo, W., Li, Z.\ H., Lu, G.\ C., Li, J.\ Q., 
Scheid, W., Lombardo, U., Schulze, H.\ J., and Shen, C.\ W.\ 
$^1$S$_0$ proton and neutron superfluidity in beta stable neutron
star matter. {\it Phys.\ Lett.\ B} {\bf 595}, 44 (2004).

\bibitem{Baldo:1998ca} Baldo, M., Elgar{\o}y, {\O}., Engvik, L., 
Hjorth-Jensen, M., and Schulze, H.-J.\ $^3$P$_2$--$^3$F$_2$ pairing
in neutron matter with modern nucleon-nucleon potentials. {\it Phys.\
Rev.\ C} {\bf 58}, 1921 (1998).

\bibitem{foulkes} Foulkes, W.\ M.\ C.,  Mitas, L., Needs, R.\ J., 
and Rajagopal, G.\ Quantum Monte Carlo simulations of solids.
\textit{Rev.\ Mod.\ Phys.} {\bf 73}, 33 (2001).

\bibitem{ceperley} Ceperley, D.\ M.\ and Alder, B.\ J.\ Ground state
of the electron gas by a stochastic method. \textit{Phys.\ Rev.\ 
Lett.} {\bf 45}, 566 (1980).

\bibitem{Pieper:2008}
Pieper, S. C.
Quantum Monte Carlo calculations of light nuclei.
\textit{Riv. Nuovo Cim.} {\bf 31}, 709 (2008).

\bibitem{AFDMC} von der Linden, W.\ A quantum Monte Carlo approach
to many-body physics. \textit{Phys.\ Rep.} {\bf 220}, 53 (1992);
Koonin, S.\ E., Dean, D.\ J., and Langanke, K.\ Shell model Monte
Carlo methods. \textit{Phys.\ Rep.} {\bf 278}, 1 (1997); 
Schmidt, K.\ E.\ and Fantoni, S.\ A quantum Monte Carlo method
for nucleon systems. \textit{Phys.\ Lett.\ B} {\bf 446}, 99 (1999).

\bibitem{HS} Stratonovich, R.\ L.\ On a method of calculating quantum 
distribution functions. \textit{Dokl.\ Akad.\ Nauk.\ SSSR} {\bf 115},
1097 (1957) [\textit{Sov.\ Phys.\ Doklady} {\bf 2}, 416 (1957)];
Hubbard, J.\ Calculation of partition functions. \textit{Phys.\ Rev.\
Lett.} {\bf 3}, 77 (1959).

\bibitem{Carlson}  Carlson, J., Chang, S.-Y., Pandharipande, V. R., and Schmidt, K. E.
Superfluid Fermi gases with large scattering length.
\textit{Phys. Rev. Lett.} {\bf 91}, 050401 (2003).

\bibitem{Fabrocini2005} Fabrocini, A. Fantoni, S., Illarionov, A. Yu., and Schmidt, K. E. 
Superfluid phase transition in neutron matter with realistic nuclear potentials and modern many-body theories.
\textit{Phys. Rev. Lett.} {\bf 95}, 192501 (2005);
Gandolfi, S., Illarionov, A. Yu., Fantoni, S., Pederiva, F.
and K. E. Schmidt, 
Equation of state of superfluid neutron matter and the calculation of the $^1$S$_0$ pairing gap.
\textit{Phys. Rev. Lett.} {\bf 101}, 132501 (2008);
Gandolfi, S., Illarionov, A. Yu., Pederiva, F., Schmidt, K. E., and Fantoni, S. 
Equation of state of low-density neutron matter, and the $^1$S$_0$ pairing gap.
\textit{Phys. Rev. C} {\bf 80}, 045802 (2009).

\bibitem{Shankar:1993pf} Shankar, R.\ Renormalization group approach
to interacting fermions. {\it Rev.\ Mod.\ Phys.} {\bf 66}, 129 (1994).

\bibitem{Schwenk:2002fq} Schwenk, A., Friman, B., and Brown, G.\ E.\
Renormalization group approach to neutron matter: Quasiparticle interactions,
superfluid gaps and the equation of state. {\it Nucl.\ Phys.\ A} {\bf 713},
191 (2003).

\bibitem{GubbelsStoof} Gubbels, K.\ B.\ and Stoof, H.\ T.\ C.\
Renormalization group theory for the imbalanced Fermi gas.
{\it Phys.\ Rev.\ Lett.} {\bf 100}, 140407 (2008).

\bibitem{Diehl} Floerchinger, S., Scherer, M., Diehl, S., and 
Wetterich, C.\ Particle-hole fluctuations in the BCS-BEC crossover,
{\it Phys.\ Rev.\ B} {\bf 78}, 174528 (2008).

\bibitem{Schwenk:2006tz} Schwenk, A.\ Superfluidity in neutron stars
and cold atoms, {\it AIP Conf.\ Proc.} {\bf 892}, 502 (2007).

\bibitem{CaoLombardoSchuck} Cao, L.\ G., Lombardo, U., and Schuck, P.\ 
Screening effects in superfluid nuclear and neutron matter within 
Brueckner theory. \textit{Phys.\ Rev.\ C} {\bf 74}, 064301 (2006).

\bibitem{Zhou} Zhou, X.-R., Schulze, H.-J., Zhao, E.-G., Pan, F., and
Draayer, J.\ P.\ Pairing gaps in neutron stars. \textit{Phys.\ Rev.\ C}
{\bf 70}, 048802 (2004).

\bibitem{Khodel:2004nt} Khodel, V.\ A., Clark, J.\ W., Takano, M., and
Zverev, M.\ V.\ Phase transitions in nucleonic matter and neutron-star
cooling, {\it Phys.\ Rev.\ Lett.} {\bf 93}, 151101 (2004).

\bibitem{Dong:2013sqa} Dong, J.\ M., Lombardo, U., and Zuo, W.\
$^3$PF$_2$ pairing in high-density neutron matter, {\it Phys.\
Rev.\ C} {\bf 87}, 062801 (2013).

\bibitem{PethickRavenhall} Pethick, C.\ J.\ and Ravenhall, D.\ G.\ 
Nuclear physics of dense matter. {\it Ann.\ N.Y.\ Acad.\ Sci.} 647, 503 (1991).

\bibitem{SchwenkFriman} Schwenk, A.\ and Friman, B.\ Polarization 
contributions to the spin dependence of the effective interaction
in neutron matter. \textit{Phys.\ Rev.\ Lett.} {\bf 92}, 082501 (2004).

\bibitem{Hammer:2012id} Hammer, H.-W., Nogga, A., and Schwenk, A.\
Three-body forces: From cold atoms to nuclei, {\it Rev.\ Mod.\ Phys.}
{\bf 85}, 197 (2013).

\bibitem{YakovlevPethick} Yakovlev D.\ G.\ and Pethick, C.\ J.\ 
Neutron star cooling. \emph{Ann.\ Rev.\ Astron.\ Astrophys.} 
\textbf{42}, 169 (2004).

\bibitem{Sjoeberg} Sj\"oberg, O.\ On the Landau effective mass in
asymmetric nuclear matter. {\it Nucl.\ Phys.\ A} {\bf 265}, 511 (1976).

\bibitem{BaldoSchulze} Baldo, M.\ and Schulze, H.-J.\  
Proton pairing in neutron stars.
\textit{Phys.\ Rev.\ C} {\bf 75}, 025802 (2007).

\bibitem{WAP91} Wambach, J., Ainsworth, T.\ L., and Pines, D.\ in
{\it Neutron Stars: Theory and Observation}. Eds. Ventura, J.\ and
Pines, D., Dordrecht: Kluwer (1991), p.~37.

\bibitem{Haensel}
Haensel, P., Potekhin, A.Y., and Yakovlev, D. G. 
{\it Neutron Stars I Equation of State and Structure}, 
Springer, New York (2007).

\bibitem{WS}  Wigner, E.\ and Seitz, F. On the constitution of
metallic sodium. \textit{Phys.\ Rev.} {\bf 43}, 804 (1933).

\bibitem{PastoreBaroniLosa} Pastore, A., Baroni, S., and Losa, C.\
Superfluid properties of the inner crust of neutron stars.
\textit{Phys.\ Rev.\ C} {\bf 84}, 065807 (2011).

\bibitem{PastoreMargueron} Pastore, A., Margueron, J., Schuck, P., 
and Vi{\~n}as, X.\ Pairing in exotic neutron-rich nuclei near the
drip line and in the crust of neutron stars. \textit{Phys.\ Rev.\ C}
{\bf 88}, 034314 (2013).

\bibitem{PastoreThermal} Pastore, A.\ Superfluid properties of the
inner crust of neutron stars.~II. Wigner-Seitz cells at finite temperature.
\textit{Phys.\ Rev.\ C} {\bf 86}, 065802 (2012).

\bibitem{cjpchamelreddy} Pethick, C.\ J., Chamel, N., and Reddy, S.\ 
Superfluid dynamics in neutron star crusts.
\textit{Prog.\ Theor.\ Phys.\ Suppl.} {\bf 186}, 9 (2010).

\bibitem{machholm} Kr\"amer, M., Pitaevskii, L., and Stringari, S.\ 
Macroscopic dynamics of a trapped Bose-Einstein condensate in the 
presence of 1D and 2D optical lattices. \textit{Phys.\ Rev.\ Lett.}
{\bf 88}, 180404 (2002); Machholm, M., Pethick, C.\ J., and Smith, H.\ 
Band structure, elementary excitations, and stability of a 
Bose-Einstein condensate in a periodic potential. \textit{Phys.\ 
Rev.\  A} {\bf 67}, 053613 (2003).

\bibitem{reddymodes} Cirigliano, V., Reddy, S., and Sharma, R.\ 
Low-energy theory for superfluid and solid matter and its application
to the neutron star crust. \textit{Phys.\ Rev.\ C} {\bf 84}, 045809 (2011).

\bibitem{chamelmodes} Chamel, N., Page, D., and Reddy, S.\ 
Low-energy collective excitations in the neutron star inner crust.
\textit{Phys.\ Rev.\ C} {\bf 87}, 035803 (2013).

\bibitem{kobyakov1}  Kobyakov, D.\ and Pethick, C.\ J.\ 
Dynamics of the inner crust of neutron stars: Hydrodynamics, elasticity,
and collective modes. \textit{Phys.\ Rev.\ C} {\bf 87}, 055803 (2013).

\bibitem{kobyakov2} Kobyakov, D.\ and Pethick, C.\ J.\ 
Towards a metallurgy of neutron star crusts.
\textit{Phys.\ Rev.\ Lett.} {\bf 112}, 112504 (2014).

\bibitem{lattimer} Lattimer, J.\ M.\ and Swesty, F.\ D.\ 
A generalized equation of state for hot, dense matter. {\it Nucl.\ 
Phys.\ A}, {\bf 535}, 331 (1991) and the website 
www.astro.sunysb.edu/dswesty/lseos.html.

\bibitem{chamel2012} Chamel, N.\ Neutron conduction in the inner 
crust of a neutron star in the framework of the band theory of solids.
\textit{Phys.\ Rev.\ C} \textbf{85}, 035801 (2012).

\bibitem{LLHydro} Landau, L.\ D.\ and Lifshitz, E.\ M.\ 
\textit{Course of Theoretical Physics, Vol.\ 6, Fluid Mechanics}, 2nd ed., 
Pergamon, Oxford (1987), p.~139.

\bibitem{chamel_haensel} Chamel, N.\ and Haensel, P.\ 
Entrainment parameters in a cold superfluid neutron star core.
\textit{Phys.\ Rev.\ C} \textbf{73}, 045802 (2006).


\end{thebibliography}
\end{document}